\documentclass[%
reprint,
twocolumn,
superscriptaddress,nofootinbib,
preprintnumbers,
nobibnotes,
 amsmath,amssymb, 
 aps, prl,
 longbibliography,
]{revtex4-1}

\usepackage{cancel}
\usepackage{accents}
\usepackage{mciteplus,slashed}
\usepackage{amssymb,cancel,amsmath}
\usepackage{dcolumn}
\usepackage{bm}
\usepackage{soul}
\usepackage[caption=false]{subfig}
\usepackage{physics}
\usepackage{booktabs}
\usepackage{comment}
\usepackage[T1]{fontenc}	
\usepackage{csvsimple}
\usepackage[bookmarksnumbered=true]{hyperref} 
\hypersetup{
    colorlinks = true,
    linkcolor = blueCERN,
    anchorcolor = blueCERN,
    citecolor = blueCERN,
    filecolor = blueCERN,
    urlcolor = blueCERN
    }

\usepackage[section]{placeins}
\usepackage[capitalise]{cleveref}
\usepackage{booktabs}
\usepackage{graphicx}
\usepackage{mathrsfs}
\graphicspath{{Figures/}}
\AtBeginDocument{
\heavyrulewidth=.08em
\lightrulewidth=.05em
\cmidrulewidth=.03em
\belowrulesep=.65ex
\belowbottomsep=0pt
\aboverulesep=.4ex
\abovetopsep=0pt
\cmidrulesep=\doublerulesep
\cmidrulekern=.5em
\defaultaddspace=.5em
}
\usepackage[dvipsnames]{xcolor}
\usepackage[normalem]{ulem}
\usepackage{fontawesome} 
\usepackage[force]{feynmp-auto}
\usepackage{bbm}
\def\iu{\mathrm{i}}
\def\e{\mathrm{e}}
\def\Meik{\mathcal{M}_{E} }

\usepackage{orcidlink}

\definecolor{amber}{rgb}{1.0, 0.75, 0.0}
\definecolor{darkturquoise}{rgb}{0.0, 0.81, 0.82}
\definecolor{mediumaquamarine}{rgb}{0.4, 0.8, 0.67}
\definecolor{coralred}{rgb}{1.0, 0.25, 0.25}
\definecolor{blueCERN}{HTML}{0033A0}
\definecolor{mygreen}{rgb}{0.0, 0.5, 0.0}

\begin{document}

\title{Partial-wave unitarity and long-range interactions}

\author{Ryan Plestid\orcidlink{0000-0003-0779-7289}}
\affiliation{Theoretical Physics Department, CERN, 1 Esplanade des Particules, CH-1211 Geneva 23, Switzerland}
\author{Pablo Qu\'ilez Lasanta\orcidlink{0000-0002-4327-2706}}
\affiliation{Theoretical Physics Department, CERN, 1 Esplanade des Particules, CH-1211 Geneva 23, Switzerland}

\date{\today}

\preprint{CERN-TH/2026-119}

\begin{abstract}
    Theories with massless particles contain $t$-channel (forward scattering) singularities that cause standard fixed order expressions for partial-wave amplitudes to be ill-defined. This presents an obstruction to systematically improvable partial-wave unitarity bounds. In this work, we study the construction of partial-wave amplitudes in a modified perturbation theory that incorporates long-range interactions focusing on the role of off-shell Coulomb modes. We find that there exists a universal description of the  forward scattering region that renders the amplitudes renormalization scale independent. The resulting partial-wave amplitudes become well defined single-scale objects without spurious dependence on the infrared regulator, and we present a practical method for their computation order-by-order in perturbation theory. 
\end{abstract}

 \maketitle

\section{Introduction} 

Unitarity places stringent constraints on the dynamics of effective field theories. For example, unitarity places bounds on the large-energy behaviour of amplitudes (the Froissart bound~\cite{Froissart:1961ux}), constrained the mass of the Higgs before its discovery (the Lee-Quigg-Thacker bound \cite{Dicus:1973gbw,Lee:1977eg,Lee:1977yc}), and limits the possible mass of elementary particle dark matter~\cite{Griest:1989wd}.  More recently, unitarity has seen a resurgence of interest with the advent of the $S$-matrix bootstrap program \cite{Paulos:2016but,Paulos:2017fhb,He:2018uxa,Cordova:2018uop,Correia:2020xtr,Kruczenski:2022lot,Correia:2025enx}. It is also a tool that can be used to constrain higher dimensional operators in the Standard Model effective field theory (SMEFT) \cite{Gounaris:1994cm,Corbett:2014ora,DiLuzio:2016sur,Corbett:2017qgl,Chang:2019vez,Remmen:2019cyz,Remmen:2020vts,Falkowski:2019tft,Remmen:2020uze,Cohen:2021gdw,Remmen:2022orj,Cao:2024vfc,Mahmud:2024iyn,Remmen:2024hry,Degrande:2025uil,Bresciani:2025toe}, and in effective field theories more generally \cite{Guerrieri:2020bto,Albert:2022oes,Albert:2023jtd,Albert:2023seb}.

Surprisingly, however, tools are relatively underdeveloped for long-range interacting theories i.e., those involving massless particles, despite our having a fairly comprehensive understanding of those theories' soft sector \cite{Yennie:1961ad,Weinberg:1965nx,Libby:1978qf,Becher:2009qa,Campiglia:2015qka,Kapec:2015ena}. In particular the long-range, $1/r$, nature of electromagnetism and gravity induces $1/t$ singularities at tree-level, and $\log(-s/t)$ singularities at loop level, that present conceptual and technical obstructions to the development of unitarity bounds \cite{Bellazzini:2021oaj,Caron-Huot:2022ugt,Caron-Huot:2022jli,Chang:2025cxc,Bellazzini:2025bay}. 

The simplest manifestation of these phenomena can be found by studying the partial-wave amplitudes defined in the azimuthally symmetric center of mass frame.  Consider the conventional (relativistic) definition of the partial-wave amplitudes $a_\ell$ for $1 + 2\rightarrow 3 + 4$ scattering of spinless particles \cite{Jacob:1959at,Itzykson:1980rh},
\begin{equation}
    \label{a_ell_def}
    a_\ell =\frac{2p_{\rm CM}}{\sqrt{s}} \frac{1}{16\pi} \int_{-1}^1 \dd\!\cos\theta ~\mathcal{M}\qty(s,t(\cos\theta)) P_\ell(\cos\theta)~. 
\end{equation}
 where $\mathcal{M}$ is the scattering amplitude, $s,t$ are the Mandelstam variables, $\theta$ is the scattering angle in the center of mass frame, and $P_\ell$ are the Legendre polynomials. Explicitly, $p_{\rm CM} = (s-m_1^2-m_2^2)\beta/(2\sqrt{s})$, the relative velocity between $1$ and $2$ is given by $\beta= \sqrt{1-m_1^2m_2^2/(p_1\cdot p_2)^2}$ and $t=-2p_{\rm CM}^2(1-\cos \theta)$.

\pagebreak 

In a short-range interacting theory  the relation between $a_\ell$ and the $S$-matrix in the $\ell^{\rm th}$ partial wave is 
\begin{equation}
    S_\ell = 1+ \iu  a_\ell ~, 
\end{equation}
Unitarity of the $S$-matrix  implies that partial-wave amplitudes satisfy,
\begin{align}
    |a_\ell -\iu |^2\leq 1~.
\end{align}
Next, let us approximate $\mathcal{M}(s,t)$ by its tree-level value. Suppose we apply \cref{a_ell_def} to a  theory with a massless photon or graviton exchanged in the $t$-channel, $\mathcal{M}(s,t)\sim \alpha/t = \alpha/[2p_{\rm CM}^2(1-\cos\theta)]$; we immediately find that $a_\ell$ is ill-defined. 
The same occurs for identical particles in the $u$-channel. 
If we include an  infrared regulator $\lambda$ (e.g., a photon mass) such that the short-range scattering theory applies we obtain $a_\ell \sim \alpha \log(\lambda/p_{\rm CM})$ such that the partial-wave cross section $\sigma_{\ell} \propto |a_\ell|^2$ is infrared divergent. Clearly the  structure of partial-wave amplitudes for long-range interacting theories requires more care than for short-range interacting theories \cite{Dollard:1964cok,Chung:1965zza,Kibble:1968sfb,Kulish:1970ut,Dollard:1971qm,Hannesdottir:2019opa,Hannesdottir:2019umk,Carney:2017jut,Carney:2017oxp,Carney:2018ygh,Lippstreu:2023vvg,Lippstreu:2025jit}. 

This is a foundational issue for partial-wave  unitarity constraints on extensions of the Standard Model. For example any SMEFT constraints derived from electromagnetically charged particles (e.g., $W^+W^- \rightarrow W^+W^-$) must contend with the fact that the renormalizable Standard Model contribution necessarily contains a $t$-channel photon contribution rendering an interpretation of $a_\ell$ ambiguous. This problem worsens at higher loop orders, where one finds infrared divergent contributions. For instance in dimensional regularization one encounters, $\sim (\alpha/\beta)\log(-t/\mu_{\rm IR}^2)$  where we emphasize that the subtraction scale $\mu_{\rm IR}$ is of infrared origin and is not removed by renormalization. Similar problems appear when trying to construct unitarity bounds for theories with charged particles and for gravity \cite{Caron-Huot:2022jli,Caron-Huot:2022ugt,Bellazzini:2021oaj,Bellazzini:2025bay}.  

As is well known, these infrared divergences are related to the inapplicability of standard perturbation theory to theories with massless particles \cite{Dollard:1964cok,Kulish:1970ut,Landau:1991wop}. This can already be seen in the non-relativistic limit where the Coulomb field, falling off like $1/r$ in $d=3$, invalidates many of the assumptions underlying conventional scattering theory. In a time-dependent formulation, the issue arises because the standard definition of the M{\o}ller operators, $\Omega_{\pm} = \lim_{\tau\rightarrow \mp \infty} \e^{\iu H \tau} \e^{-\iu H_0 \tau}$ does not converge. When real photons (or gravitons) are included one must furthermore contend with on-shell soft radiation. 

In this paper we study the modified time-dependent perturbation theory due to Dollard that {\it does} have a convergent $\tau\rightarrow \mp \infty$ limit. Using techniques from effective field theory we show how to relate amplitudes computed naively using dimensional regularization to the Dollard-derived results. We then study the limit $|t| \ll p_{\rm CM}^2$ that is the source of forward scattering singularities discussed above. We find that the theory can be, and must be, resumed to all orders in perturbation theory prior to decomposing in partial waves. This sharpens a recent conjecture~\cite{FuentesZamoro:2025exp} (motivated by non-relativistic theory) into a constructive demonstration that $t$-channel singularities do not prevent the definition of a finite and unitary partial-wave  $S$-matrix, $S_\ell$, for relativistic field theories with massless particles. 

\section{Long-range scattering theory} \label{Sec:LongRangeScatteringTh}

Standard scattering theory presumes an interaction which falls off faster than $1/r$ as $r\rightarrow \infty$. For Coulomb-like potentials, all complications arise from pair-wise interactions between the particles. Therefore, for simplicity, in what follows we focus on the elastic scattering of two particles $p_1 p_2 \rightarrow p_1' p_2'$ with charges $z_1$ and $z_2$. As is well known, the so-called ``Coulomb parameter'', $\eta =z_1 z_2\,\alpha/\beta$  (when not written we take $z_1=z_2=1$) naturally appears in the analysis, and we find it convenient to use the Mandelstam variable $t=(p_1-p_1')^2=(p_2-p_2')^2$. Although we focus on $2\rightarrow 2$ scattering the generalization to more than one particle simply involves summing over all pair-wise Coulomb interactions. 

Let us now review the standard construction of time-dependent scattering theory (see e.g., Refs.~\cite{taylor2012scattering,Weinberg:1995mt}) The $S$-matrix is defined as $S_{ab}\equiv \langle {\psi^{(-)}_b}|{\psi^{(+)}_a}\rangle$ where $|{\psi^{(\pm)}}\rangle$ are in- and out-states (i.e., exact eigenstates of the Hamiltonian $H$). The in- and out-states can be constructed via M{\o}ller operators,\!\footnote{These identities only need to hold when smeared against a smooth wavepacket; we ignore this technical detail.} 
\begin{equation}
    |\psi^{(\pm)}_a\rangle = \Omega^{(\pm)}\ket{a} = \lim_{\tau\rightarrow \mp \infty}  \e^{\iu H \tau} \e^{-\iu H_0 \tau}\ket{a}~, 
\end{equation}
where $\ket{a}$ is an eigenstate of $H_0$ (e.g., a plane wave). The M{\o}ller operators exist in a short-range theory because as $\tau\rightarrow \infty$, and particles separate from one another asymptotically, the free evolution cancels between the free and full Hamiltonians. 

This convergence is spoiled when there is a massless particle that mediates an unscreened Coulomb-like potential in the theory. This can be easily understood semi-classically by imagining the time-integral of the potential energy $\int \dd t ~ \alpha/|\vb{x}_0 - \vb{v} t| \sim \alpha/|\vb{v}| \log \tau/T_0$. Dollard showed that this can be fixed by appending an additional operator that acts on $\ket{a}$ and that scales like $\log \tau$ such that modified ``Dollard operator'' \cite{Dollard:1964cok,Dollard:1971qm}
\begin{equation}
    \Omega^{(\pm)}_D\ket{a} = \lim_{\tau\rightarrow \mp \infty}  \e^{\iu H \tau} \e^{-\iu H_0 \tau} \e^{-\iu \eta ~{\rm sgn}(\tau) \log\qty(\frac{|\tau|}{T_0})}\ket{a}
\end{equation}
exists where $T_0$ is a function of $a$ and we have anticipated the relativistic velocity $\beta$ in place of $|\vb{v}|$. These ideas can be extended to include real photons \cite{Chung:1965zza,Kulish:1970ut}, or other asymptotic reference dynamics \cite{Hannesdottir:2019opa,Hannesdottir:2019umk}, however we focus here on the infrared divergence that arises from off-shell photons (the Coulomb field). The Dollard operator then serves as the basis for a finite and convergent perturbation theory. 

The constant $T_0(\ket{a})$ must be fixed to supply unambiguous amplitudes. Since $T_0$ is defined in the soft/eikonal limit, $|t|\ll s$,
it should be some function of $m_1,m_2$ and $v_{1\mu}$ and $v_{2\mu}$ (the four-velocities which act as conserved labels in the eikonal limit). Furthermore, the result should be symmetric under $\{m_1,v_1\} \leftrightarrow \{m_2,v_2\}$. 

We can uniquely fix $T_0$ in two cases where exact results are known; the non-relativistic $\beta\rightarrow 0$ limit, and the static limit $m_1/m_2 \rightarrow 0$. In both limits the problem is soluble because the two-body problem reduces to a single particle scattering off a
static Coulomb potential.  Therefore, the simplest matching computation is to compare the in-state wavefunction in coordinate space,
 \begin{equation}
     \psi^{(+)}_{\vb{k}}(\vb{x}) = \mel{\vb{x}}{\Omega^{(+)}_D}{\vb{k}}~,
 \end{equation}
where $\vb{x}$ and $\vb{k}$ are the position and momentum three-vectors of the particle.
We may choose $1/|\vb{x}| \ll \vb{k}$, while we must have that $T \gg \vb{x}$ by assumption. Since the scales are well separated, the perturbative series is well suited to a method of regions analysis \cite{Beneke:1997zp}. 

First consider the matrix element of the M{\o}ller operator at fixed time $T$ which is equivalent to studying $U_I(-T,0)=\e^{-\iu H T}\e^{-\iu H_0 T}$ in for $T \gg \vb{x}$ with $U_I$ the interaction picture evolution operator. Using the Dyson-series for $U_I(-T,0)$ we can work order-by order in perturbation theory
\begin{align}
    \!\!\!\!\! \mel{\vb{x}}{U_I(-T,0)}{\vb{k}} = \e^{\iu \vb{k}\cdot\vb{x}} &+ \int_{-T}^{0} \dd \tau \mel{\vb{x}}{ \e^{\iu H_0 \tau} V \e^{-\iu H_0 \tau}}{\vb{k}} \nonumber \\
    &\hspace{0.3\linewidth}+~~\ldots~
\end{align}
Let us begin at $O(\alpha)$ where upon inserting a complete set of momentum eigenstates, $\ketbra{\vb{L}+\vb{k}}$, with integration measure $(\dd L)\equiv \dd^3L/(2\pi)^3$,  we find
\begin{equation}
   \int (\dd L) \e^{\iu (\vb{k}+\vb{L})\cdot\vb{x}} \int_{-T}^{0} \dd \tau \mel{\vb{k}+\vb{L}}{ \e^{\iu H_0 \tau} V \e^{-\iu H_0 \tau}}{\vb{k}}~.
   \label{Eq:IntegralOalpha}
\end{equation}
One can now explicitly see how the finite time, $T$, acts as an infrared regulator. Acting with $H_0$ on the plane wave states and carrying out the time integral one finds, 
\begin{equation}
  \int_{-T}^{0} \dd \tau ~ \e^{\iu(E_{\vb{k}+\vb{L}} - E_{\vb{k}}) \tau} = \frac{ 1- \e^{-\iu(E_{\vb{k}+\vb{L}} - E_{\vb{k}}) T}}{\iu (E_{\vb{k}+\vb{L}} - E_{\vb{k}})} ~.
  \label{Eq:T}
\end{equation}
In conventional calculations, where $T\rightarrow \infty(1 -\iu 0)$ with $\iu 0$ the causal regulator, the second term in the numerator would be dropped. With $T$ held finite, as $E_{\vb{k}+\vb{L}}\rightarrow E_{\vb{k}}$ the expression remains finite and logarithmic infrared divergences are effectively ``cutoff'' by the scale $T$. 

We may now apply the method of regions \cite{Beneke:1997zp} (see \cref{App:IntegralsWaveMatching}). Consider the hierarchy $T\gg |\vb{x}| \gg 1/|\vb{k}|$. The ``hard'' region where $|\vb{L}| \sim |\vb{k}|$ is suppressed by the rapidly oscillating exponentials, and can be neglected. The ``soft'' region where $|\vb{L}| \sim 1/|\vb{x}|$ gives the standard infrared divergent result, 
\begin{equation}
    \tilde{\psi}^{(1)}_{\rm soft}(\vb{x},T) = \e^{\iu \vb{k}\cdot \vb{x}} \times  e^2 \int (\dd L) \frac{1}{\vb{L}^2} \frac{1}{\vb{v} \cdot \vb{L} }  \e^{\iu \vb{L}\cdot \vb{x}}~,
\end{equation}
where $|\vb{v}|=\beta$. The  asymptotic region gives, 
\begin{equation}
    \tilde{\psi}^{(1)}_{\rm asym.}(\vb{x},T)=  \e^{\iu \vb{k}\cdot \vb{x}}  e^2 \int (\dd L) \frac{1}{\vb{L}^2} \frac{1-\e^{-\iu \vb{v} \cdot \vb{L} T}}{\vb{v} \cdot \vb{L} }~, 
\end{equation}
which has an ultraviolet divergence. 
Both integrals can be evaluated using dimensional regularization and added together. Their $1/\epsilon$ poles cancel and their sum $\tilde{\psi}^{(1)}_{\rm soft}(\vb{x},T)+ \tilde{\psi}^{(1)}_{\rm asym.}(\vb{x},T)$ yields,
\begin{equation}
     \tilde{\psi}^{(1)}(\vb{x},T) =  \e^{\iu \vb{k}\cdot \vb{x}}  ~\iu \eta  \qty[ \log( \frac{r-z }{2 \beta T} )]~.
\end{equation}
This is the correct large-$r$ behavior for a Coulomb wavefunction (at first order in $\alpha$), and comes with the advertised $\log(T)$ dependence.

Finally, it is easy to show that this result exponentiates. Both the soft- and asymptotic regions have eikonal Feynman rules.  For example at second order one finds
\begin{equation}
    \begin{split}
    \!\!\!e^4&\int (\dd L_1) (\dd L_2)\!\frac{\e^{\iu (\vb{L}_1+\vb{L}_2)\cdot \vb{x}}}{\vb{L}_1^2 \vb{L}_2^2} 
    \int_{-T}^0 \dd \tau \!\int_{-T}^\tau \dd \tau' \e^{\iu \vb{v}\cdot \vb{L}_1 \tau} \e^{\iu \vb{v}\cdot \vb{L}_2 \tau'}\!\!\! \\
    &=~\frac{1}{2!}\qty( e^2 \int (\dd L_1) \!\frac{\e^{\iu \vb{L}_1\cdot \vb{x}}}{\vb{L}_1^2 } 
    \int_{-T}^0 \dd \tau  ~\e^{\iu \vb{v}\cdot \vb{L}_1 \tau} )^2~,
    \end{split}
\end{equation}
which follows by a simple change of variables. The pattern persists to higher orders and the resummation of the full series leads to an exponentiated one-loop result, 
\begin{equation}
     \tilde{\psi}^{(+)}(\vb{x},T)=  \e^{\iu \vb{k}\cdot \vb{x}}  \times \exp\qty{ \iu \eta \qty[ \log( \frac{r-z }{2 \beta T} )]}~.
\end{equation}
Finally to obtain the actual wavefunction (that does not depend on $T$) we must append the Dollard phase  since we are actually interested in the matrix element of $\Omega_D^{(+)}$ rather than $\Omega^{(+)}$. Defining $\psi_{\vb{k}}^{(+)}(\vb{x}) = \tilde{\psi}^{(+)}(\vb{x},T)\e^{\iu \eta \log(T/T_0)}$ we then find
\begin{equation}
    \psi_{\vb{k}}^{(+)}(\vb{x})=\e^{\iu \vb{k}\cdot \vb{x}}  \times \exp\qty{ \iu \eta  \qty[ \log( \frac{r-z }{2 \beta T_0} )]}
\end{equation}
We may now fix $T_0$ if we have exact solutions in hand, which essentially amounts to a matching calculation for the Dollard operator. 

In the non-relativistic limit the relevant logarithm is $\log(2p_{\rm CM, NR} r(1-\cos\theta))$ \cite{messiah2014quantum} where $p_{\rm CM, NR}= \mu v$ with $\mu$ the reduced mass and $v$ the non-relativistic relative velocity. For relativistic kinematics exact results are known in the static limit where the mass of one of the particles tends to infinity and acts as a static Coulomb potential. The answer in the relativistic case is $2 |\vb{p}| r(1-\cos\theta)$ \cite{rose1961relativistic} when evaluated in the rest frame of the heavy particle. Choosing $T_0(\ket{a}) = 1/(2 \beta p_{\rm CM})$ reproduces both of these two limits where $p_{\rm CM}= (s-m_1^2-m_2^2)\beta/(2\sqrt{s})$ (as above). For kinematics away from the static and non-relativistic limit this remains a (physically well motivated) ansatz. 

Although we have demonstrated this behavior for the wavefunction, the same Dollard phase must properly dress {\it all } scattering theory quantities. Indeed, having fixed $T_0$, the fate of any infrared divergences related to the Coulomb phase are now fixed. We will see in what follows that the forward scattering amplitude computed with $T_0$ properly reproduces the known exact results from non-relativistic theory. 

To sum up, accounting for the long-range Coulomb dynamics requires a modified perturbation theory. Infrared divergences in the naive amplitude resum into a phase and when evaluated at finite times lead to terms that scale as $\log(T)$. This logarithmic dependence on time cancels against the Dollard phase. For matrix elements, one effectively takes an infrared divergent amplitude $ \mathcal{M}_\epsilon$ (computed for example in dimensional regularization) and trades it for a $T$-dependent amplitude $\mathcal{M}_\epsilon\rightarrow \widetilde{\mathcal{M}}(T)$. This $T$-dependence is then canceled by the Dollard phase, 
\begin{equation}
    \begin{split}
    \mathcal{M}&=\widetilde{\mathcal{M}}(T)\, \e^{\iu [ \phi_D(-T)-\phi_D(T)]}\\
    &=\widetilde{\mathcal{M}}(T)\, \e^{\iu \eta \log(4 \beta^2 T^2 p_{\rm CM}^2)} \,,
    \end{split}
\end{equation}
where 
\begin{equation}
     \phi_D(\mathcal{T}) = -\eta~ {\rm sgn}(\mathcal{T})
  \log\!\big(2\beta\, p_{\rm CM}\, |\mathcal{T}|\big) ~,
\end{equation}
and the final result, $\mathcal{M}$, is $T$ independent. The matching calculation described above illustrates how to relate infrared divergent\footnote{Care must be taken to separate IR from UV logarithms. The replacement rule, $\mu_{\rm IR}^2 \rightarrow |t|_{\rm max}$, applies only to IR divergences.} amplitudes computed in dimensional regularization to those obtained in the Dollard perturbation theory (see also Refs.~\cite{Hannesdottir:2019opa,Hannesdottir:2019umk} and \cref{app:litt-comp}).

\section{Forward scattering} \label{Sec:ForwardScattering}
Let us now consider fixed-$t$ kinematics where $|t| \ll p_{\rm CM}^2$. This is the portion of the integral over $\dd\!\cos\theta$ where forward scattering singularities present technical difficulties. One may imagine splitting $\int_{-1}^1 \dd\!\cos\theta  = \int_{-1}^{1-\epsilon} \dd\!\cos\theta + \int_{1-\epsilon}^{1} \dd\!\cos\theta$, and the following discussion applies to the near-forward region $\int_{1-\epsilon}^{1} \dd\!\cos\theta$. 

When $|t| \ll p_{\rm CM}^2$ one can define an effective field theory that separates photons with virtuality  $|L^2|\sim |t|$ (which appear dynamically in the theory), from short-distance contributions from photons of virtuality  $|L^2|\sim p_{\rm CM}^2$ (which are captured in Wilson coefficients); here $L$ denotes a loop momentum. If the theory contains other massive states (e.g., a virtual $Z$-boson) that mediate scattering these contributions are also captured in short-distance Wilson coefficients. 

In the effective theory the Feynman rules for massive particles\footnote{The treatment of forward scattering for massless particles is discussed in e.g., Ref.~\cite{Rothstein:2016bsq,Rothstein:2024nlq}.} are eikonal (see e.g., Ref.~\cite{Manohar:2000dt}): propagators go like $\iu /(v\cdot L)$ and interaction vertices with photons are given by $(\iu Q e)v_\mu$ where $v_\mu$ is the four-velocity of the particle. In what follows we find it convenient to keep masses (replacing $v_\mu \rightarrow 2 m v_\mu$ in the above formulae) to match onto the relativistic normalization conventions used in \cref{a_ell_def} i.e., we  define the eikonal amplitude $\Meik$ to have the same mass dimensions and normalization as $\mathcal{M}$. One can easily check that when both particles $1$ and $2$ carry electric charge, the dominant contribution to the amplitude scales as $1/|t|$ and that this observation holds at arbitrary loop order. Terms that are less singular at small-$|t|$ correspond to higher dimensional operators in the effective theory.

Let us denote amplitudes computed in the eikonal theory by $\Meik$ and those in the full theory by $\mathcal{M}$. We will denote perturbative orders in $\alpha$ with a superscript and the power-expansion in the EFT by subscripts. To leading order one has, 
\begin{equation}
    \label{EFT-exp}
    \mathcal{M}= \Meik + O(\sqrt{|t|}/p_{\rm CM})~. 
\end{equation}
The leading-power amplitude $\Meik$ (i.e., the eikonal approximation) comes from graphs in the effective theory where only photons are exchanged. As mentioned above, we express $\Meik$ in terms of a finite-time matrix element, and the ``Dollard phase'', 
\begin{equation}
    \Meik= \widetilde{\mathcal M}_E(T) \e^{\iu [\phi_D(-T)-\phi_D(T)]}~, 
\end{equation}
As shown in \cref{App:EikonalAmp}, $\widetilde{\mathcal{M}}_E(T)$ is given by,
\begin{equation}    \label{eikonal-finite-time}
   \!\!\! \widetilde{\mathcal{M}}_{E}(T)=  \frac{16\pi\alpha (p_1\cdot p_2)}{t} \e^{-\iu \eta \log(|t|\beta^2 T^2)} \frac{\Gamma(1+\iu \eta)}{\Gamma(1-\iu \eta)}  ~. 
\end{equation}
where $p_1\cdot p_2=(s-t-u)/4$.
Combining with the Dollard phase then gives (recall $|t|_{\rm max}=4p_{\rm CM}^2$),
\begin{equation}
    \label{eikonal-post-dollard}
   \!\! \Meik = -16\pi\alpha\qty(\frac{p_1\cdot p_2}{-t} )\e^{-\iu \eta \log(|t|/|t|_{\rm max})} \frac{\Gamma(1+\iu \eta)}{\Gamma(1-\iu \eta)}~. 
\end{equation}
This resolves the conceptual issue related to an infrared subtraction point, $\mu_{\rm IR}$, discussed in the introduction. Unlike in QCD where $\mu_{\rm IR}$ is related to the scale dependence of a long-distance matrix element (the parton distribution function), here it is a symptom of an ill-defined perturbation theory (for plane-wave states in vacuum). With the inclusion of the Dollard phase it is ``lifted'' to the scale $|t|_{\rm max}$ and  reproduces the conjectured prescription in Ref.~\cite{FuentesZamoro:2025exp} using $-\iu \eta \log (|t|/|t|_{\rm max})=-\iu \eta \log (\frac{1-\cos\theta}{2})$.

Let us remark that this construction also explains why the eikonal approximation is exact for the non-relativistic Coulomb amplitude (a well known but curious fact). Scattering solutions to the Schr\"odginer equation at different energies are related to one another by a rescaling of the coordinate $\vb{x}$ with $\eta$ held fixed. Since the scattering amplitude $f(\theta)$ is obtained from the $|\vb{x}|\rightarrow \infty$ asymptotics (with $\theta$ held fixed) so cannot depend explicitly on $\vb{p}$ when written in terms of $\eta$: the momentum transfer $Q^2=-t$ is then the only dimensionful scale and dimensional analysis demands $\mathcal{M}= Q^{-2} g(\eta, \log(Q^2))$ for some function $g$. This in turn implies that all power corrections in \cref{EFT-exp} must vanish. The problem then effectively becomes single-scale and must be given by the eikonal amplitude to all orders in $t/p_{\rm CM}^2$. 

The above procedure can be used to constrain the amplitude in the $|t|\rightarrow 0$ limit for any theory. In more general theories that do not have special conformal properties one finds power corrections. As we will now see, control over the $|t|\rightarrow 0$ limit is sufficient to control the partial-wave amplitudes and to obtain sensibly unitary results. 

\section{Unitarity: $S$-matrix vs. $T$-matrix}
Let us now compute the re-scaled Legendre transform of $\Meik(\cos\theta)$, which we define by, cf. \cref{a_ell_def}, 
\begin{equation}
\!\!\!\!
\mathcal{J}_{\ell} \{F\} \equiv \frac{2p_{\rm CM}}{\sqrt{s}}\frac{1}{16\pi} \int_{-1}^1 \dd\!\cos\theta ~F(\cos\theta) P_\ell(\cos\theta)~.
\end{equation}
We define the $S$-matrix by $S_\ell=1+\iu a_\ell = \e^{2\iu \delta_\ell}$ with $\delta_\ell$ the phase-shift of the $\ell^{\rm th}$ partial wave. Naively we expect $\mathcal{J}_\ell\{\mathcal{M}_E\}$ to give us something that is proportional to $a_\ell$, the analog of the $T$-matrix in the $\ell^{\rm th}$ partial-wave  channel.  As we will see, this would lead to a paradox since the Legendre transform of $\Meik$ is $O(1)$ as $\alpha\rightarrow0$ as was recently pointed out in Ref.~\cite{FuentesZamoro:2025exp}.  

Let us begin with the $\ell=0$ Legendre transform of $\Meik$. Here we obtain, 
\begin{align}
   \mathcal{J}_{\ell=0}\{\mathcal{M}_E\}
   &=-\frac{\eta}{2}\int_{-1}^1 \dd\!\cos\theta \Big(\frac{2}{1-\cos\theta}\Big)^{1-\iu \eta}~\frac{\Gamma(1+\iu \eta)}{\Gamma(1-\iu \eta)} \nonumber\\
    &= -\iu~\frac{\Gamma(1+\iu \eta)}{\Gamma(1-\iu \eta)}~. 
\end{align}
Notice that as $\eta\rightarrow 0$ this answer tends to $-\iu$ and so it is natural to study  the Legendre transform of $\mathcal{J}_\ell\{+\iu \mathcal{M}_E\}$ which then gives the non-relativistic $s$-wave phase shift~\cite{messiah2014quantum,gottfried2003quantum,schiff1955quantum,FuentesZamoro:2025exp}. More generally one finds that
\begin{align}
    \mathcal{J}_\ell\{\iu \mathcal{M}_E\}
    &= ~\frac{\Gamma(1+\ell+\iu \eta)}{\Gamma(1+\ell-\iu \eta)}~.
\end{align}
It is now clear the the partial-wave amplitudes so-obtained are $S_\ell^{E}$, i.e., unitary phases,  rather than $a_\ell$. This occurs despite the fact that $\mathcal{M}_E$ begins at $O(\alpha)$ because the Legendre transform is regulated by the exponent $\iu \eta$, and therefore introduces a power of $1/\alpha$ \cite{FuentesZamoro:2025exp}.

The eikonal approximation is universal in the small-$|t|$ limit (it applies to relativistic and non-relativistic kinematics and does not depend on spin). We therefore conclude that the partial-wave amplitudes of all theories with long-range interactions should be identified with $S_\ell$ rather than $a_\ell$. This agrees with similar observations made in Refs.~\cite{Herbst1974,Landau:1991wop,Lin:2000tin,Lippstreu:2023vvg}.

Notice that this conclusion is fundamentally different than what one would obtain from regulating the theory with a photon mass. In that case one would identify the Legendre transform of $\mathcal{M}(\lambda,\cos\theta)$ with $a_\ell(\lambda)$ which would then diverge like $\log(\lambda)$ as $\lambda\rightarrow 0$. Indeed, one finds for the exactly solvable (in the $s$-wave) non-relativistic Hluthen potential that $S_\ell = S_\ell^E\e^{\iu \delta_H(\lambda)}$ with $\delta_H(\lambda)$ logarithmically divergent \cite{HulthenLaurikainen1951,Ma:1954}. Although $S_\ell^E\e^{\iu \delta_H}$ is still unitary, when perturbative corrections are added interference terms sensitive to $\log(\lambda)$ can potentially spoil the infrared safety of partial-wave  unitarity; one must properly handle the long-range Coulomb dynamics. It is interesting that partial-wave amplitudes provide a counter example to the folk theorem that the Coulomb phase always disappears when considering squared amplitudes (it does not for $|a_\ell|^2$). 

One may interpret this observation by thinking in impact parameter space $\dd \cos\theta \rightarrow \dd^2 b$. The integral extends to impact parameters $b \gg 1/\lambda$ such that partial-wave amplitudes probe arbitrarily long-distances. One therefore obtains different answers if $\lambda$ is sent to zero before or after defining the partial-wave amplitudes. This corresponds to different choices of infrared physics. For example studying the theory in vacuum or in the presence of a medium which screens the long-range Coulomb field (note this is conceivable for electromagnetism but not gravity).  Most unitarity bounds are derived considering $2\rightarrow 2$ scattering in vacuum. This choice corresponds to setting $\lambda=0$ before defining the $S$-matrix and therefore necessitates the modified scattering theory discussed above.

\section{A subtraction scheme}
The above discussion makes it clear that the small-|$t$| region is amenable to all-orders resummation. One can then slice the integration regime into $\cos\theta\in[-1,1-\epsilon]$ and $\cos\theta\in[1-\epsilon,1]$, do a fixed order treatment on the first domain, a resummed treatment on the second, and then match-and-merge the two answers at the boundary. Although this is conceptually straightforward it seems an unwieldy approach. 

Instead, we suggest the following subtraction scheme. Consider the amplitude $\mathcal{M}(s,t)$ computed in Dollard's modified perturbation theory. Write 
\begin{align}
    S_\ell = &\frac{2p_{\rm CM}}{\sqrt{s}} \frac{1}{16\pi}\bigg\{\int_{-1}^1 \dd\!\cos\theta P_\ell(\cos\theta)~ \iu \Meik \nonumber\\
    &\hspace{0.2\linewidth}+\iu \int_{-1}^1 \dd\!\cos\theta P_\ell(\cos\theta) \qty[\mathcal{M} - \Meik]  \bigg\}~.
    \nonumber
    \\
    &=S_\ell^E +\iu\hspace{0.001\linewidth}a_\ell'~,
\end{align}
 where $S_\ell^E = \Gamma(1+\ell+\iu \eta)/\Gamma(1+\ell-\iu \eta)$. 
The integral involving $ \qty[ \mathcal{M} - \Meik]$ is free from non-integrable forward scattering singularities. This behavior holds for arbitrary spin, composite or elementary, particles in both the relativistic and non-relativistic regimes because the eikonal approximation is universal.  

The quantity $\qty[ \mathcal{M} - \Meik]$ can therefore be computed and integrated order-by-order in perturbation theory.   Any logarithmic infrared divergences $\log(-t/\mu_{\rm IR})$  are shifted to $\log(-t/|t|_{\rm max})$ via the Dollard construction sketched above. Terms that scale as $\log^n(-t/|t|_{\rm max})$ are integrable. The Legendre transform therefore introduces no powers of $1/\alpha$. If $u$-channel singularities are present  (as is the case for identical particles) then a separate eikonal treatment applies to the $\cos\theta\rightarrow -1$ region. 

The relevant quantity of interest is then,
\begin{equation}
    a_\ell' =  \frac{2p_{\rm CM}}{\sqrt{s}} \frac{1}{16\pi} \int_{-1}^1 \dd\!\cos\theta P_\ell(\cos\theta) \qty[ \mathcal{M} - \Meik]~. 
\end{equation}
Partial wave unitarity becomes $|S_\ell^E + \iu a_\ell'|\leq 1$, or 
\begin{equation}
      |a_\ell' |^2 \leq 2 {\rm Im}\qty( a_\ell'^* S_\ell^E)~. 
\end{equation}
The perturbative expansion of $a_\ell'$ can be combined with the expansion of $S_\ell^E$ to study perturbative unitarity bounds order-by-order in $\alpha$ unencumbered by infrared divergences. Real photon emission presents separate conceptual issues that are orthogonal to the $t$-channel singularities that are the focus of this work; we comment briefly on these issues in our conclusions. 

To make the discussion concrete let us consider the scattering of a high-energy  electron\footnote{For spin-$1/2$ particles one must use Jacob-Wick helicity amplitudes \cite{Jacob:1959at} (or an equivalent formalism) for $J=\ell+1/2$. These details do not affect the forward scattering region. } in a Coulomb field. To retain the relativistic conventions for scattering amplitudes, we will keep the heavy-particles current, $2M v_\mu$, explicit. In the rest-frame of the static potential the tree-level amplitude is \cite{Dalitz:1951ah,Hill:2012rh} (dropping $m/E$ corrections), 
\begin{equation}
    \begin{split}
    \mathcal{M}^{(0)} &= \frac{4\pi \alpha}{Q^2} \bar{u}(p+q) \gamma^\mu u(p) (2Mv_\mu)  \\
    &=\frac{4\pi \alpha}{Q^2} \qty(4 M E \cos\qty(\tfrac{\theta}{2})) 
    \end{split}
\end{equation}
where $Q^2 =-t= 2E^2(1-\cos\theta)$ with $E$ the energy of the electron, and we have explicitly evaluated the Dirac spinors. At one-loop we have \cite{Dalitz:1951ah,Hill:2012rh}
\begin{equation}
\begin{split}
    \mathcal{M}^{(1)} = \alpha \mathcal{M}^{(0)}   \bigg[ &\frac{\pi}{2}\frac{Q}{2E+Q} + \iu \log(\frac{Q^2}{4E^2}) \\
    &~~~~~~~~~
    + \iu \frac{Q^2}{4E^2-Q^2} \log\qty(\frac{Q}{2E})   \bigg]  ~, 
\end{split}
\end{equation}
where we have regulated the infrared divergence with dimensional regularization, and then replaced the infrared scale with $Q^2_{\rm max}=4E^2$ via the Dollard phase. 

We now subtract from these expressions\footnote{Recall that we use a consistent state normalization between $\mathcal{M}$ and $\Meik$, which differs from common EFT conventions \cite{Manohar:2000dt}.}
\begin{align}
    \Meik^{(0)} = \frac{4\pi\alpha}{Q^2} (4 M E) 
\end{align}
and 
\begin{align}
    \Meik^{(1)} = \Meik^{(0)} \times \qty[ \iu \alpha \log(\frac{Q^2}{4E^2})] ~. 
\end{align}
This then gives
\begin{equation}
\begin{split}
    \!\!\!\!\!\qty[ \mathcal{M} - \Meik]&=\frac{4\pi\alpha}{Q^2} (4 M E)\bigg[ \qty(1-\cos(\tfrac{\theta}{2}))   \\
    &+ \iu \alpha \qty(1-\cos(\tfrac{\theta}{2}))\log\qty( \frac{Q^2}{4E^2})\\
    &+ \alpha\frac{\pi}{2}\frac{Q}{2E+Q} + \iu \alpha \frac{Q^2}{4E^2-Q^2} \log\qty(\frac{Q}{2E})   \bigg]~\\
    &+~  O(\alpha^2)~.
\end{split}
\end{equation}
It is easy to check that when computing $a'_{\ell}$ this gives finite integrals order-by-order in $\alpha$.  For example $(1-\cos(\theta/2))/Q^2 \sim O(1)$ as $\cos\theta \rightarrow 1$. 

When vertex corrections are included, as opposed to box-like topologies discussed above,  infrared divergences will appear in $[\mathcal{M}-\Meik]$; these are related to infrared divergences from real photon emission. Real-photons present separate technical difficulties \cite{Kulish:1970ut} since they introduce new inelastic channels. By including modified asymptotic dynamics that include both the Coulomb field (as is done by the Dollard operator) and real photons, a convergent and unitary \cite{Carney:2017jut,Carney:2017oxp,Carney:2018ygh} $S$-matrix may be defined, however it is dependent on an infrared cutoff used to define the dressing.  This is necessary because real photons above the cut  are tracked as separate inelastic channels.

For recent discussion of the application of these ideas to the $S$-matrix bootstrap program see Refs.~\cite{Bellazzini:2021oaj,Caron-Huot:2022ugt,Caron-Huot:2022jli,Lippstreu:2023vvg,Chang:2025cxc,Bellazzini:2025bay,Lippstreu:2025jit}.  We stress that the Coulomb singularity is distinct from real photons and is not sensitive to how the Hilbert space is partitioned into ``soft'' and ``hard'' sectors. 

\section{Conclusions}
Theories involving photons and gravitons contain forward scattering divergences that cause partial-wave amplitudes to be infrared sensitive. These issues are related to the presence of long-range Coulomb interactions which do not asymptotically decouple at long-distances. In this work we have revisited the modifications necessary to obtain an infrared finite perturbation theory, with an emphasis on the relation between partial-wave amplitudes and infrared divergences in dimensionally regulated scattering amplitudes.

One of the key deliverables of our analysis is the connection between finite time and dimensional regularization as an infrared regulator (see \cite{Hannesdottir:2019opa,Hannesdottir:2019umk} for a closely related perspective). This correspondence is enabled by a method-of-regions analysis, and allows one to perform practical computations using dimensional regularization and then to trade dependence on the infrared subtraction scale for late-time logarithms. These same logarithms are removed after the application of the Dollard phase, yielding single-scale partial-wave amplitudes. 

The presence of long-range interactions modifies how one interprets the Legendre transform of the scattering amplitude. In a short range theory this yields $a_\ell$ (where $S_\ell= 1+\iu a_\ell$) whereas when there is a $|t|$-channel singularity one instead finds that the Legendre transform of the leading eikonal amplitude yields $S_\ell^E$, while the pieces that are finite as $|t|\rightarrow 0$ may be identified with a modified $a_\ell'$ such that $S_\ell=S_\ell^E +\iu a_\ell'$ is unitary.

In summary we have provided a constructive connection between dimensionally regulated amplitudes and the modified Dollard perturbation theory. This resolves infrared ambiguities related to $t$-channel singularities. We have presented a subtraction scheme for practical computations that allows an order-by-order computation of $a_\ell'$; this can be directly applied to SMEFT amplitudes {\it including} contributions from the Standard Model photon. The properties of Legendre transformed amplitudes imply that one must carefully interpret partial-wave coefficients in the presence of long range interactions. We hope this work is helpful both for SMEFT practitioners, and for the $S$-matrix bootstrap program.

\pagebreak

\textbf{Acknowledgments:} We thank Cliff Burgess, Ben Grinstein, Dimitrios Kosmopoulos,  Julio Parra-Martinez,  Michael Saavedra,  Frank Tackmann, Gherardo Vita,   and Alexander Zhiboedov for helpful discussions. This work began at the annual CERN theory retreat 2025, and we thank the organizers for the event.  P.Q. acknowledges support by the European Union's Horizon 2020 research and innovation programme under the Marie Sk\l odowska-Curie Postdoctoral Fellowship grant agreement No 101207780 - AxionCount.

\appendix
\crefalias{section}{appendix}
\section{Partial wave conventions}
Let us briefly review the connection between relativistic and non-relativistic partial-wave  scattering conventions. We begin with the non-relativistic construction, and then consider scalar-QED as a relativistic example. 

In the non-relativistic context one typically studies the scattering amplitude $f(\theta)$ which has units of length or inverse momentum. The differential cross section is expressed as $\dd \sigma/\dd \Omega=|f(\theta)|^2$. When the potential is short-ranged the partial-wave amplitudes, defined by $S_\ell= 1+\iu a_\ell$, are obtained by a Legendre transform, 
\begin{equation}
    a_\ell =   \int_{-1}^1 \dd c_\theta~ P_\ell(c_\theta)   k f(\theta)~, 
    \label{Eq:PartialWaveApp}
\end{equation}
which then implies that 
\begin{equation}
    f(\theta) = \frac{1}{2k} \sum_\ell  (2\ell+1) P_\ell(c_\theta) a_\ell ~.
\end{equation}

For the non-relativistic Coulomb problem, $f(\theta)$ is given by, 
\begin{equation}
     \!\!\!\!\!f_C(\theta) = -\frac{\eta}{k(1-\cos\theta)} \e^{-\iu \eta \log\qty(\frac{1-\cos\theta}{2}) } \frac{\Gamma(1+\iu \eta)}{\Gamma(1-\iu\eta)}~.
\end{equation}
Taking its Legendre transform yields $S_\ell =\Gamma(1+\ell+\iu \eta)/\Gamma(1+\ell-\iu \eta)$ rather than $a_\ell$. The breakdown of standard expectations from short-range scattering theory can be traced to the fact that the large-$r$ asymptotics of the Coulomb wavefunction are a series in $1/(r-z)= 1/r \times 1/\cos\theta$ rather than $1/r$.   

Next, consider the tree-level scattering of two distinguishable charged relativistic scalars $\phi(p_1) \Phi(p_2) \rightarrow \phi(p_3) \Phi(p_4)$,
\begin{equation}
    \mathcal{M}^{(0)} = e^2 \frac{(p_1+p_3)\cdot (p_2+p_4) }{t}  = e^2 \qty(\frac{s-u}{t}) ~. 
\end{equation}
Next, using $s+t+u = 2 m_1^2+2m_2^2$ one obtains, 
\begin{equation}
    \mathcal{M}^{(0)} = e^2 \qty(\frac{2s-2m_1^2-2m_2^2 + t}{t}) ~.
\end{equation}
The $+t$ in the numerator maps onto a contact operator in the effective theory describing forward scattering. Using the relativistic conventions we include a factor of $2p_{\rm CM}/\sqrt{s} \times (1/16\pi)$. One can easily show that 
\begin{equation}
    \frac{2p_{\rm CM}}{16\pi\sqrt{s}}  \times e^2\frac{(2s-2m_1^2-2m_2^2)}{t}
    = \frac{\eta}{(1-\cos\theta)} ~.
\end{equation}
We thus find that $\frac{2p_{\rm CM}}{16\pi\sqrt{s}}\mathcal{M}^{(0)}_{\rm pole}$ matches only $k f^{(0)}(\theta)$ in the non-relativistic conventions, where $\mathcal{M}_{\rm pole}$ is the residue of the $t$-channel pole. This also shows how the correct normalization (proportional to $\eta$) is obtained in the relativistic result; this is necessary for a unitary $S$-matrix upon accounting for loop-corrections. One is thus lead to (as given above) 
\begin{equation}
    \label{a_ell_def-repeated}
    a_\ell =\frac{2p_{\rm CM}}{\sqrt{s}} \frac{1}{16\pi} \int_{-1}^1 \dd c_\theta ~\mathcal{M}\qty(s,t(c_\theta)) P_\ell(c_\theta)~,
\end{equation}
for relativistic scalar scattering in a short-range interacting theory.

Partial-wave amplitudes are more complicated when incident or outgoing particles have spin. The partial-wave expansion then involves Wigner $d$-functions instead of the Legendre polynomials (Jacob-Wick helicity formalism~\cite{Jacob:1959at}) due to spin-orbit coupling. When using the result of Dalitz~\cite{Dalitz:1951ah} we require partial-wave amplitudes for spin-$1/2$ electrons. In fact, we only need the $J=\tfrac12$, $\ell=0$, partial-wave
\begin{equation}
    \hspace{-0.015\linewidth}a_{\ell=0,J=1/2} =  \frac{2p_{\rm CM}}{\sqrt{s}} \frac{1}{16\pi} \int_{-1}^1 \dd c_\theta \cos\tfrac{\theta}{2} ~ \mathcal{M}\qty(s,t(c_\theta))~.
    \label{Eq:PartialWaveAppFermion}
\end{equation}
One may find general formulae in Refs.~\cite{Jacob:1959at,ParticleDataGroup:2024cfk}.

\section{Photon mass regulator}
We can better understand the interplay of infrared divergences and partial-wave  unitarity by studying amplitudes computed with a photon mass as an infrared regulator. In particular we specialize to amplitudes computed in spin-1/2 QED where one-loop results, due to Dalitz \cite{Dalitz:1951ah}, are known for a static Yukawa potential i.e., with a finite photon mass $\lambda$. The results are given in terms of complicated integrals $I$ and $J$ for arbitrary values of $Q$, $E$, $m$, and $\lambda$.

Let us take the external fermions to be massless ($m=0$, $E=p=|\mathbf{p}|$) and  expand the Dalitz result~\cite{Dalitz:1951ah} in the small~$\lambda$ limit (which is singular and contains $\log(\lambda)$ terms). In order to obtain finite partial-wave amplitudes the tree-level propagator $1/(Q^2+\lambda^2)$ must be retained.\!\footnote{Using the full $\lambda$-dependent expressions from Ref.~\cite{Dalitz:1951ah} we have verified that \cref{a-ell-0-dalitz,a-ell-1-dalitz} are the correct small-$\lambda$ asymptotics for the partial-wave amplitudes. } The tree-level (Born) amplitude reads 
\begin{equation}\label{eq:fD0}
  \mathcal{M}^{(0)}(Q,\lambda)
  = \frac{4\pi \alpha\, (4 M E)\cos\theta/2}{Q^2+\lambda^2}\,,
\end{equation}
where $Q^2\equiv 2E^2(1{-}\cos\theta)$. 
At one loop the result is complicated, however the following small-$\lambda$ approximant is sufficient to track the $\log^2\lambda$ and $\log\lambda$ dependence of the partial-wave amplitudes,\!\footnote{See the discussion beneath \cref{eq:small-lambda-region-limit}. } 
\begin{multline}\label{eq:fD1}
  \mathcal{M}(Q,\lambda)
  \simeq \mathcal{M}^{(0)}(Q,\lambda)
  \bigg[1+ 
   \alpha
  \bigg(\frac{\pi}{2}\frac{Q}{2E+Q}\\ 
  +\iu\left(\log\frac{Q^2}{\lambda^2}
      + \frac{Q^2}{4E^2-Q^2} \log\frac{Q}{2E}\right)\bigg)
  \bigg]\,.
\end{multline}

The tree-level $s$-wave ($J=\ell+1/2$ with $\ell=0$) amplitude is obtained using \cref{Eq:PartialWaveAppFermion} and is given by
 \begin{align}
        \label{a-ell-0-dalitz}a^{(0)}_{\ell=0}(\lambda)=-\alpha\left[1+2\log\frac{\lambda}{2E}\right]~.
 \end{align}
At one loop, using \cref{eq:fD1}, we control only the $\log(\lambda)$-enhnaced terms and so drop ${\cal O}(1)$ contributions, 
  \begin{multline}
  \label{a-ell-1-dalitz}
  a_{\ell=0}^{(1)}(\lambda)= \alpha^2  \times
 \bigg\{\qty[-\pi \log\frac{\lambda}{2E} + {\cal O}(1)] \\ 
   +\,\iu\Big[  2\log \frac{\lambda}{2E} \qty(1+\log \frac{\lambda}{2E}) + {\cal O}(1) \Big]\bigg\}~.
\end{multline}

The logarithmic pieces in the second-order partial-wave amplitude are a requirement of perturbative unitarity. The tree-level amplitude is real and $\sim \alpha \log(\lambda/2E)$. The unitarity of the $S$-matrix thus requires an imaginary part at one-loop order that scales as $\iu \alpha^2 \log^2(\lambda/2E)$. One can check explicitly that to $O(\alpha^2)$ partial-wave unitarity is respected by \cref{a-ell-0-dalitz,a-ell-1-dalitz} up to ${\cal O}(\alpha^2)$ finite terms i.e., those not enhanced by $\log(\lambda)$. 

If one wishes to control the finite pieces in the partial-wave amplitudes, more care is required. This is because terms that are suppressed at fixed-$Q$ (like $\lambda^2/(Q^2+\lambda^2)^2$) can supply ${\cal O}(1)$ contributions to the partial-wave amplitudes. These come entirely from the region of integration where $Q^2\sim \lambda^2$ or equivalently where $1-\cos\theta \sim \lambda^2 /E^2$. For example, if one takes the full expressions from \cite{Dalitz:1951ah} and scales $Q \rightarrow \varepsilon Q$ and $\lambda\rightarrow \varepsilon \lambda$, then expands in $\varepsilon$ the expression is 
\begin{equation}
\begin{split}
    \label{eq:small-lambda-region-limit}
    \lim_{\varepsilon\rightarrow 0} & ~\varepsilon^2 \frac{\mathcal{M}^{(1)}(\varepsilon \lambda, \varepsilon Q)}{4 M E \cos\theta/2 } =  4\pi \alpha^2 \iu \frac{ \log \left(\frac{\sqrt{4 \lambda ^2+Q^2}+Q}{\sqrt{4 \lambda ^2+Q^2}-Q}\right)}{Q \sqrt{4 \lambda ^2+Q^2}}~,
\end{split}
\end{equation}
where the $\varepsilon^2$ is motivated by $\dd \cos\theta = \dd Q^2 /(2E^2)\sim \varepsilon^2$. One clearly sees that the $Q\sim \lambda$ region requires keeping the full $\lambda$ dependence in the above expression. The full $\lambda$-dependence is {\it not} retained in \cref{eq:fD1} e.g., the above logarithm would turn into $\log(Q/\lambda)$. This is why ${\cal O}(\alpha^2)$ contributions (without $\log\lambda$ enhancement) cannot be reliably tracked using \cref{eq:fD1}. This exercise also demonstrates the infrared sensitivity of partial-wave amplitudes. As is well appreciated in other contexts, it is much more difficult to calculate with a photon mass than without one.

No smooth $\lambda\to 0$ limit exists at any fixed order in the expansion of $a_\ell(\lambda)$. Partial wave amplitudes defined in vacuum and obtained with a photon mass are therefore regulator dependent (even when unitarity is fully respected).
The solution to this pathology is connected to the Dollard construction in \cref{Sec:LongRangeScatteringTh,Sec:ForwardScattering}: the IR-divergent piece $\sim\log(\lambda^2)$ is precisely what the Dollard phase absorbs, replacing $\lambda^2\to |t|_{\rm max}=4p_{\rm CM}^2$ and making the partial waves finite and essentially single-scale objects.

Having studied a relativistic example at second-order in perturbation theory, we turn to the non-relativistic regime where exact results can be obtained. 
Consider the Hulth\'{e}n potential~\cite{Hulthen1942,HulthenLaurikainen1951,Ma:1954},
\begin{equation}\label{eq:Hulthen}
  V(r) = -\alpha \,\frac{\lambda\, e^{-\lambda r}}{1-e^{-\lambda r}}\,,
\end{equation}
where $\lambda$, the analog of the photon mass, acts as an inverse screening length.
This potential has two key properties.
First, it interpolates between a Coulomb potential at short distances, $V(r)\to -V_0/r$ for $r\ll 1/\lambda$, and an exponentially screened potential $V(r)\sim -V_0\lambda\, e^{-\lambda r}$ at large~$r$.
Second, the Schr\"{o}dinger equation for this potential can be solved exactly for the $s$-wave ($\ell=0$).
The exact $s$-wave phase shift is known in closed form~\cite{Hulthen1942,Ma:1954},
\begin{equation}
    S_{\ell=0} = \frac{\Gamma(1+\iu a)}{\Gamma(1-\iu a)} \times  \frac{\Gamma(1+\iu b)\Gamma(1+\iu c)}{\Gamma(1-\iu b)\Gamma(1-\iu c)} ~,
\end{equation}
where
\begin{align}
    a &= \frac{\iu p}{\lambda} \qty[ 1-\qty(1-\frac{2\alpha m \lambda}{p^2})^{1/2}] ~,\\
    b &= \frac{\iu p}{\lambda} \qty[ 1+\qty(1-\frac{2\alpha m \lambda}{p^2})^{1/2}] ~,\\
    c &= \frac{2 p}{\lambda}~.
\end{align}
In the small-$\lambda$ limit one can readily check that 
\begin{equation}
    S_{\ell=0}\rightarrow  \frac{\Gamma(1+\iu \eta )}{\Gamma(1-\iu \eta)} \times  \e^{\iu \eta \log\qty(\frac{4p^2}{\lambda^2})}  + O(\lambda/p)~.
\end{equation}
We see that the $s$-wave scattering matrix contains an additional infrared divergent phase. It then follows that the $s$-wave's scattering amplitude, $a_0 = (S_0-1)/\iu$ scales as $a_0\sim \alpha \log(p/\lambda)$ in agreement with perturbation theory. This exactly solvable example makes it clear that the introduction of a screening parameter changes the structure of the partial-wave amplitudes, and affects the mapping between fixed-angle amplitudes and the partial-wave  $S$-matrix.

\section{Eikonal amplitude at finite time}
\label{App:EikonalAmp}
Let us discuss the computation of $\widetilde{\mathcal{M}}_E$ from \cref{eikonal-finite-time}. We desire the analog of the $S$-matrix at finite time, which is given by the transition amplitude
\begin{equation}
    \mathcal{S}_{fi}(T) = \mel{f}{U_I(-T,T)}{i}~.
\end{equation}
As is well known, because the kinetic term in the eikonal theory is convective, the interaction picture evolution operator exponentiates in coordinate space. To obtain $\tilde{\mathcal{M}}_E(T)$ we should compute $\mathcal{S}_{fi}(T)- \braket{f}{i}$ but working at $|t|\neq 0$ ensures that the forward scattering piece vanishes. In the eikonal limit the momentum transfer is spatial and transverse to the beam $t=-|\vb{Q}_\perp|^2$, and one finds,
\begin{equation}
     \!\!
    \iu \widetilde{\mathcal {M}}_E(T)\sim  \int \dd^2b~ \e^{\iu \vb{Q}_\perp \cdot \vb{b}} \exp\qty[\iu \int_{-\beta T}^{\beta T} \frac{-\eta }{\sqrt{z^2+b^2}} \dd z ]~, \!
\end{equation}
where we have omitted the normalization.

Next use 
\begin{equation}
    \int_{-\beta T}^{\beta T} \frac{\eta }{\sqrt{z^2+b^2}} \dd z = 2 \eta  \log \left(\frac{\sqrt{b^2+(\beta T)^2}+\beta T}{\sqrt{b^2+(\beta T)^2}-\beta T}\right)~.
\end{equation}
For $1/T \ll |\vb{Q}_\perp|$ only the $b \ll \beta T$ region contributes to the integral, and we obtain 
\begin{equation}
     \!\!
    \iu \widetilde{\mathcal {M}}_E(T)\sim  \int \dd^2b~ \e^{\iu \vb{Q}_\perp \cdot \vb{b}} \qty(\frac{4\beta^2 T^2}{b^2})^{-\iu \eta} ~,
\end{equation}
which then yields, after restoring the relativistic conventions used herein, 
\begin{equation}
    \begin{split}
    \iu \widetilde{\mathcal {M}}_E(T) = - \iu &\frac{16\pi\alpha (p_1\cdot p_2)}{Q_\perp^2}\\
    &\hspace{0.1\linewidth}\times \e^{-\iu \eta \log(Q_\perp^2 \beta^2 T^2)} \frac{\Gamma(1+\iu \eta)}{\Gamma(1-\iu \eta)} ~,
    \end{split}
\end{equation}
which agrees with \cref{eikonal-finite-time} upon identifying $t=-Q_\perp^2$.
%

\section{Integrals for wavefunction matching}
\label{App:IntegralsWaveMatching}
In this appendix, we carry out a matching calculation that fixes the reference scale
$T_0$ in the Dollard operator introduced in \cref{Sec:LongRangeScatteringTh}. We evaluate
the $O(\alpha)$ distortion of the in-state wavefunction in the far field and compare it to
the known exact Coulomb wavefunction, and thus we obtain $T_0$.
Consider the integral in \cref{Eq:IntegralOalpha}
\begin{equation}
    I = \frac{e^2}{\beta} \int (\dd L) \frac{1}{\vb{L}^2} \frac{1-\e^{-\iu \vb{v} \cdot \vb{L} T}}{\vb{u} \cdot \vb{L} }  \e^{\iu \vb{L}\cdot \vb{x}}~.
\end{equation}
Consider the hierarchy $T\gg |\vb{x}| \gg |\vb{k}|$. The ``hard'' region where $\vb{L} \sim \vb{k}$ is suppressed by the rapidly oscillating exponentials, and can be neglected. The ``soft'' region where $\vb{L} \sim \vb{x}$ gives the standard infrared divergent result, 

Using the method of regions \cite{Beneke:1997zp,Jantzen:2011nz} for the hierarchy $T\gg |\vb{x}| \gg |\vb{k}|$, we split this into two pieces, $I=I_{\rm asym.} + I_{\rm soft}$, 
\begin{align}
    I_{\rm asym.} &= e^2\int (\dd L) \frac{1}{\vb{L}^2} \frac{1-\e^{-\iu \vb{v} \cdot \vb{L} T}}{\vb{v} \cdot \vb{L} } ~,\\
    I_{\rm soft} &= e^2 \int (\dd L) \frac{1}{\vb{L}^2} \frac{1}{\vb{v} \cdot \vb{L} }  \e^{\iu \vb{L}\cdot \vb{x}}~,
\end{align}
where $|\vb{v}|=\beta$. We will now proceed to evaluate these contributions with dimensional regularization. Using a Schwinger parameter on the linear propagator, we obtain the following representation 
\begin{equation}
     I_{\rm soft} = e^2  \int \dd s \int (\dd L) \frac{1}{\vb{L}^2}  \e^{\iu \vb{L}\cdot (\vb{x}-\vb{v}s) }~. 
\end{equation}
Evaluating the $d$-dimensional Fourier transform then yields, 
\begin{equation}
    I_{\rm soft} = \iu \frac{e^2}{\beta}  \frac{\Gamma(\tfrac12 -\epsilon)}{(4\pi)^{3/2-\epsilon}} \int_0^\infty  \dd s \qty(\frac{1}{(\vb{x} - \vb{u}s)^2})^{1/2-\epsilon} ~. 
\end{equation}
where $\vb{u}=(\vb*{0},1)$ is a unit vector in the $\vb{v}$ direction. It is convenient to now use coordinates $\vb{x}=(\vb{b},z)$  and obtain, 
\begin{equation}
    I_{\rm soft} = \frac{\iu e^2}{\beta}   \frac{\Gamma(\tfrac12 -\epsilon)}{(4\pi)^{3/2-\epsilon}} \int_{-z}^\infty  \dd s \qty(\frac{1}{s^2 + b^2})^{1/2-\epsilon} ~. 
\end{equation}
This integral can now be split into two pieces 
\begin{equation}
    \begin{split}
    \!\!\!I_{\rm soft} = \frac{\iu e^2}{\beta}  \frac{\Gamma(\tfrac12 -\epsilon)}{(4\pi)^{3/2-\epsilon}} 
    \bigg\{& \int_{0}^\infty  \dd s \qty(\frac{1}{s^2 + b^2})^{1/2-\epsilon} \\
    &+ \int_{-z}^0  \dd s \qty(\frac{1}{s^2 + b^2})^{1/2-\epsilon}  \bigg\} ~. 
    \end{split}
\end{equation}
If we now restore the coupling $e^2$ and re-write the expressions in terms of the $\overline{\rm MS}$ coupling using
\begin{equation}
    e^2=\frac{(4\pi)^{1+\epsilon}}{\Gamma(1+\epsilon)} \mu^{-2\epsilon} \bar{\alpha}  ~, 
\end{equation}
we obtain
\begin{equation}
\begin{split}
    I_{\rm soft} &= \frac{\iu \bar{\alpha}}{\beta} \frac{(4\pi)^{1+\epsilon}}{\Gamma(1+\epsilon)} \frac{\Gamma(\tfrac12 -\epsilon)}{(4\pi)^{3/2-\epsilon}}\qty(\frac{b^2}{\mu^2})^\epsilon    
     \\
    &\times\bigg\{ \frac{ \sqrt{\pi } \Gamma \left(-\epsilon\right)}{2 \Gamma (\tfrac12-\epsilon )} + \int_{-z/b}^0  \dd s \qty(\frac{1}{s^2 + 1})^{1/2-\epsilon}  \bigg\} ~.
    \end{split}
\end{equation}
We can evaluate the second integral inside the curly braces at $\epsilon=0$ and expand in $\epsilon$ to obtain, 
\begin{align}
    I_{\rm soft} 
    &= \frac{\iu \bar{\alpha}}{\beta} \qty[ -\frac{1}{4\epsilon} - \frac12 \log (4\pi \e^{\gamma_{\rm E}} \frac{b}{\mu}) - \frac12 \log(\frac{r-z }{b}) ]~ \nonumber \\
    &=\frac{\iu \bar{\alpha}}{\beta} \qty[ -\frac{1}{4\epsilon} - \frac12 \log (4\pi \e^{\gamma_{\rm E}} \frac{r-z }{\mu})  ]~.
\end{align}

The same steps can be carried out in the asymptotic region. Dropping scaleless integrals, one finds that 
\begin{align}
    I_{\rm asym.}  &= \frac{\iu \bar{\alpha}}{\beta} \frac{(4\pi)^{1+\epsilon}}{\Gamma(1+\epsilon)} \frac{\Gamma(\tfrac12 -\epsilon)}{(4\pi)^{3/2-\epsilon}}     
    \qty{ - \int_0^\infty \dd s \frac{1}{(s + \beta T)^{1-2\epsilon}  } }\nonumber  \\
    &= \frac{\iu \bar{\alpha}}{\beta} \frac{(4\pi)^{1+\epsilon}}{\Gamma(1+\epsilon)} \frac{\Gamma(\tfrac12 -\epsilon)}{(4\pi)^{3/2-\epsilon}}     
    \qty{ - \frac{(\beta T)^{2\epsilon}}{2\epsilon} } \\
    &= \frac{\iu \bar{\alpha}}{\beta} \qty[ \frac{1}{4\epsilon} + \frac12 \log (4\pi \e^{\gamma_{\rm E}} \frac{2\beta T}{\mu}) ]\nonumber 
\end{align}
Adding the two expressions together we arrive at 
\begin{equation}
    I = \iu \eta  \qty[ \log( \frac{r-z }{2 \beta T} )]
\end{equation}
where we have used $\bar{\alpha}|_{\epsilon=0}= \alpha=e^2/(4\pi)$. 

\section{Comparison to the literature \label{app:litt-comp}}
Infrared divergences and the Coulomb phase are a very old problem that is well studied in  non-relativistic quantum mechanics, nuclear physics, high-energy scattering, and in the context of soft theorems. In order to orient what we have presented in context we therefore review other treatments in the literature.

\subsection*{The non-relativistic Coulomb problem}
The non-relativistic Coulomb problem is an illuminating examples because exact solutions can be obtained with multiple different methods \cite{messiah2014quantum,Landau:1991wop}. Partial wave phase shifts can be obtained directly by working in spherical coordinates, or indirectly by working in parabolic-cylinder coordinates, obtaining $f(\theta)$, and computing Legendre transforms. Furthermore, since much of our intuition in scattering theory stems from non-relativistic potential scattering, the non-relativistic limit serves as a useful laboratory where we can explicitly see how our standard assumptions break down in the presence of a Coulomb potential. 

Most of the features we have commented on above appear already in this problem. For instance the Legendre transform of $f(\theta)$ yields $S_\ell$ rather than $a_\ell$ \cite{Landau:1991wop}. The ability to construct scattering-state wavefunctions directly provides further insight into this peculiar property. Typically (i.e., in short-range scattering problems) one uses the asymptotic form $\psi(r)\sim \e^{\iu k z} + f(\theta)\e^{\iu k r}/r$. Working at fixed-$r$ a Legendre transform can be performed and $f(\theta)$ related directly to $a_\ell$. 

In the Coulomb problem, however, one has large-$r$ asymptotics governed by $1/k(r-z) = 1/(k r[1-\cos\theta])$. When performing the Legendre transform, these singular $1/(1-\cos\theta)$ terms spoil the fixed-$r$ analysis. When one attempts a partial-wave decomposition at large-$r$ the $1/(1-\cos\theta)$ singularities render the angular integrals ill defined, and the large-$r$ asymptotics become unreliable; one must instead perform the partial-wave  decomposition first, and then the large-$r$ limit. Indeed one can readily check that the separation into an incident plane wave and outgoing spherical wave fails for this reason. These issues are discussed in detail in the following textbooks \cite{Landau:1991wop,messiah2014quantum}.

\subsection*{Distorted-wave perturbation theory}
It is well known that if Coulomb effects are dealt with at leading-order then a modified perturbation theory may be constructed. This is often known as the distorted-wave Born series \cite{weinberg2013lectures}, or the Furry-picture \cite{Furry:1951bef,schweber2005introduction}; it is essentially a time-independent perturbation theory using the Coulomb Hamiltonian as the unperturbed basis. Its application is limited to problems where the Coulomb field can be treated as a static classical background field. 

Recently, Lippstreu has written two papers proposing a variant of distorted wave perturbation theory \cite{Lippstreu:2023vvg,Lippstreu:2025jit}. Formal properties of the $S$-matrix are studied therein, and a variety of explicit computations carried out. As in the modified Dollard perturbation theory, the results are infrared finite, however because the strict eikonal approximations is not used,  calculations are substantially more complicated than those presented in this work. 

There is a small technical difference between Lippstreu's results and our own. We have provided an ansatz for $T_0$ constrained by the requirements of symmetry under $(m_1,v_1)\leftrightarrow (m_2,v_2)$, and explicit results in the static and non-relativistic limits. Lippstreu works in the rest frame of one of the particles, and properly reproduces the static limit $m_1/m_2\rightarrow 0$, however the symmetry property and non-relativistic limit are not properly reproduced. This is because $p_{\rm CM} \neq p_{1,\rm rest}^2$ where $p_{1,\rm rest}$ is the momentum of particle-one in the rest frame of particle-two. In the non-relativistic limit one can deduce unambiguously that $p_{\rm CM}$ is the correct scale.

\subsection*{Modified time-dependent perturbation theory}
The approach followed in this work is to revisit a time-dependent perturbation theory with modified asymptotic dynamics. We have effectively used the original proposal due to Dollard, but generalizations to include real photon emissions were developed by Fadaeev and Kulish, and later studied by many authors. 

The implementation of the Dollard operator is discussed in detail in the main text so we do not belabor this issue. Infrared divergences related to real photons are handled in a conceptually similar manner by including eikonal couplings to the dynamical photon field \cite{Kulish:1970ut}. In Coulomb gauge, the Dollard operator arises from the longitudinal modes, while infrared divergences related to on-shell modes correspond to transverse polarizations. 

We expect that in order to obtain a unitary $S$-matrix, it is necessary to include asymptotic reference dynamics like the canonical Fadaeev-Kulish dressing. In particular it has been explicitly demonstrated that without this dressing the off-diagonal entries of the $S$-matrix decohere, and one is left with an explicitly non-unitary operator \cite{Carney:2017jut,Carney:2017oxp,Carney:2018ygh}. For practical applications, an infrared dressing scheme introduces an infrared regulator $\mathcal{E}$ that separates ``soft'' photons, from ``hard'' photons where the latter modes appear as dynamical inelastic final states in the theory. This is qualitatively different from the $1/t$ Coulomb-like singularities, which need no such separation (they are entirely an off-shell effect). 

In a practical implementation, we therefore expect the following to occur: An infrared regulator scheme is chosen for the asymptotic dynamics. If one computes exclusive processes e.g., scattering with a vanishing or fixed number of hard photons, then the $S$-matrix will be regulator dependent and will satisfy unitarity bounds with strict inequality since there are always inelastic final states available. The incoherent sum of $S$ matrix elements will, of course, be infrared-safe (independent of the $\mathcal{E}$) for sufficiently inclusive observables. 

A distinction between much of the literature on modified reference dynamics and our work is the focus on practical computation. In particular we have shown how to append the appropriate phase onto amplitudes in which infrared divergences have been regulated by analytically continuing the space-time dimension. This is very similar to the work by Schwartz and Hannesdottir \cite{Hannesdottir:2019opa,Hannesdottir:2019umk} where different reference dynamics based on modern effective field theories are considered, and the much more difficult non-Abelian case confronted. 

There is minor technical difference between our work and Refs.~\cite{Hannesdottir:2019opa,Hannesdottir:2019umk}, because we explicitly retain a finite time $T$ in our calculations, which introduces a new region. After combining the relevant regions we find that the expressions have a well defined limit as $T\rightarrow \infty$ such that the final answer is $T$-independent. Hannesdottir and Schwartz use propagators which implicitly assume that the large-$T$ limit exists in all regions (i.e, terms like $\exp[-\iu E(T-\iu 0)]$ are dropped everywhere). The ``asymptotic'' region, with loop momenta scaling as $1/T$, can introduce new finite pieces in the computation, and so one should, in general, check that the two methods agree. It is possible that these finite pieces always cancel between the two regions, but this remains to be explicitly demonstrated (if it is indeed true). 

\vfill
\pagebreak

\bibliography{biblio.bib}

\begin{thebibliography}{81}%
\makeatletter
\providecommand \@ifxundefined [1]{%
 \@ifx{#1\undefined}
}%
\providecommand \@ifnum [1]{%
 \ifnum #1\expandafter \@firstoftwo
 \else \expandafter \@secondoftwo
 \fi
}%
\providecommand \@ifx [1]{%
 \ifx #1\expandafter \@firstoftwo
 \else \expandafter \@secondoftwo
 \fi
}%
\providecommand \natexlab [1]{#1}%
\providecommand \enquote  [1]{``#1''}%
\providecommand \bibnamefont  [1]{#1}%
\providecommand \bibfnamefont [1]{#1}%
\providecommand \citenamefont [1]{#1}%
\providecommand \href@noop [0]{\@secondoftwo}%
\providecommand \href [0]{\begingroup \@sanitize@url \@href}%
\providecommand \@href[1]{\@@startlink{#1}\@@href}%
\providecommand \@@href[1]{\endgroup#1\@@endlink}%
\providecommand \@sanitize@url [0]{\catcode `\\12\catcode `\$12\catcode
  `\&12\catcode `\#12\catcode `\^12\catcode `\_12\catcode `\%12\relax}%
\providecommand \@@startlink[1]{}%
\providecommand \@@endlink[0]{}%
\providecommand \url  [0]{\begingroup\@sanitize@url \@url }%
\providecommand \@url [1]{\endgroup\@href {#1}{\urlprefix }}%
\providecommand \urlprefix  [0]{URL }%
\providecommand \Eprint [0]{\href }%
\providecommand \doibase [0]{http://dx.doi.org/}%
\providecommand \selectlanguage [0]{\@gobble}%
\providecommand \bibinfo  [0]{\@secondoftwo}%
\providecommand \bibfield  [0]{\@secondoftwo}%
\providecommand \translation [1]{[#1]}%
\providecommand \BibitemOpen [0]{}%
\providecommand \bibitemStop [0]{}%
\providecommand \bibitemNoStop [0]{.\EOS\space}%
\providecommand \EOS [0]{\spacefactor3000\relax}%
\providecommand \BibitemShut  [1]{\csname bibitem#1\endcsname}%
\let\auto@bib@innerbib\@empty
\bibitem [{\citenamefont {Froissart}(1961)}]{Froissart:1961ux}%
  \BibitemOpen
  \bibfield  {author} {\bibinfo {author} {\bibfnamefont {Marcel}\ \bibnamefont
  {Froissart}},\ }\bibfield  {title} {\enquote {\bibinfo {title} {{Asymptotic
  behavior and subtractions in the Mandelstam representation}},}\ }\href
  {\doibase 10.1103/PhysRev.123.1053} {\bibfield  {journal} {\bibinfo
  {journal} {Phys. Rev.}\ }\textbf {\bibinfo {volume} {123}},\ \bibinfo {pages}
  {1053--1057} (\bibinfo {year} {1961})}\BibitemShut {NoStop}%
\bibitem [{\citenamefont {Dicus}\ and\ \citenamefont
  {Mathur}(1973)}]{Dicus:1973gbw}%
  \BibitemOpen
  \bibfield  {author} {\bibinfo {author} {\bibfnamefont {D.~A.}\ \bibnamefont
  {Dicus}}\ and\ \bibinfo {author} {\bibfnamefont {V.~S.}\ \bibnamefont
  {Mathur}},\ }\bibfield  {title} {\enquote {\bibinfo {title} {{Upper bounds on
  the values of masses in unified gauge theories}},}\ }\href {\doibase
  10.1103/PhysRevD.7.3111} {\bibfield  {journal} {\bibinfo  {journal} {Phys.
  Rev. D}\ }\textbf {\bibinfo {volume} {7}},\ \bibinfo {pages} {3111--3114}
  (\bibinfo {year} {1973})}\BibitemShut {NoStop}%
\bibitem [{\citenamefont {Lee}\ \emph {et~al.}(1977{\natexlab{a}})\citenamefont
  {Lee}, \citenamefont {Quigg},\ and\ \citenamefont {Thacker}}]{Lee:1977eg}%
  \BibitemOpen
  \bibfield  {author} {\bibinfo {author} {\bibfnamefont {Benjamin~W.}\
  \bibnamefont {Lee}}, \bibinfo {author} {\bibfnamefont {C.}~\bibnamefont
  {Quigg}}, \ and\ \bibinfo {author} {\bibfnamefont {H.~B.}\ \bibnamefont
  {Thacker}},\ }\bibfield  {title} {\enquote {\bibinfo {title} {{Weak
  Interactions at Very High-Energies: The Role of the Higgs Boson Mass}},}\
  }\href {\doibase 10.1103/PhysRevD.16.1519} {\bibfield  {journal} {\bibinfo
  {journal} {Phys. Rev. D}\ }\textbf {\bibinfo {volume} {16}},\ \bibinfo
  {pages} {1519} (\bibinfo {year} {1977}{\natexlab{a}})}\BibitemShut {NoStop}%
\bibitem [{\citenamefont {Lee}\ \emph {et~al.}(1977{\natexlab{b}})\citenamefont
  {Lee}, \citenamefont {Quigg},\ and\ \citenamefont {Thacker}}]{Lee:1977yc}%
  \BibitemOpen
  \bibfield  {author} {\bibinfo {author} {\bibfnamefont {Benjamin~W.}\
  \bibnamefont {Lee}}, \bibinfo {author} {\bibfnamefont {C.}~\bibnamefont
  {Quigg}}, \ and\ \bibinfo {author} {\bibfnamefont {H.~B.}\ \bibnamefont
  {Thacker}},\ }\bibfield  {title} {\enquote {\bibinfo {title} {{The Strength
  of Weak Interactions at Very High-Energies and the Higgs Boson Mass}},}\
  }\href {\doibase 10.1103/PhysRevLett.38.883} {\bibfield  {journal} {\bibinfo
  {journal} {Phys. Rev. Lett.}\ }\textbf {\bibinfo {volume} {38}},\ \bibinfo
  {pages} {883--885} (\bibinfo {year} {1977}{\natexlab{b}})}\BibitemShut
  {NoStop}%
\bibitem [{\citenamefont {Griest}\ and\ \citenamefont
  {Kamionkowski}(1990)}]{Griest:1989wd}%
  \BibitemOpen
  \bibfield  {author} {\bibinfo {author} {\bibfnamefont {Kim}\ \bibnamefont
  {Griest}}\ and\ \bibinfo {author} {\bibfnamefont {Marc}\ \bibnamefont
  {Kamionkowski}},\ }\bibfield  {title} {\enquote {\bibinfo {title} {{Unitarity
  Limits on the Mass and Radius of Dark Matter Particles}},}\ }\href {\doibase
  10.1103/PhysRevLett.64.615} {\bibfield  {journal} {\bibinfo  {journal} {Phys.
  Rev. Lett.}\ }\textbf {\bibinfo {volume} {64}},\ \bibinfo {pages} {615}
  (\bibinfo {year} {1990})}\BibitemShut {NoStop}%
\bibitem [{\citenamefont {Paulos}\ \emph {et~al.}(2017)\citenamefont {Paulos},
  \citenamefont {Penedones}, \citenamefont {Toledo}, \citenamefont {van Rees},\
  and\ \citenamefont {Vieira}}]{Paulos:2016but}%
  \BibitemOpen
  \bibfield  {author} {\bibinfo {author} {\bibfnamefont {Miguel~F.}\
  \bibnamefont {Paulos}}, \bibinfo {author} {\bibfnamefont {Joao}\ \bibnamefont
  {Penedones}}, \bibinfo {author} {\bibfnamefont {Jonathan}\ \bibnamefont
  {Toledo}}, \bibinfo {author} {\bibfnamefont {Balt~C.}\ \bibnamefont {van
  Rees}}, \ and\ \bibinfo {author} {\bibfnamefont {Pedro}\ \bibnamefont
  {Vieira}},\ }\bibfield  {title} {\enquote {\bibinfo {title} {{The S-matrix
  bootstrap II: two dimensional amplitudes}},}\ }\href {\doibase
  10.1007/JHEP11(2017)143} {\bibfield  {journal} {\bibinfo  {journal} {JHEP}\
  }\textbf {\bibinfo {volume} {11}},\ \bibinfo {pages} {143} (\bibinfo {year}
  {2017})},\ \Eprint {http://arxiv.org/abs/1607.06110} {arXiv:1607.06110
  [hep-th]} \BibitemShut {NoStop}%
\bibitem [{\citenamefont {Paulos}\ \emph {et~al.}(2019)\citenamefont {Paulos},
  \citenamefont {Penedones}, \citenamefont {Toledo}, \citenamefont {van Rees},\
  and\ \citenamefont {Vieira}}]{Paulos:2017fhb}%
  \BibitemOpen
  \bibfield  {author} {\bibinfo {author} {\bibfnamefont {Miguel~F.}\
  \bibnamefont {Paulos}}, \bibinfo {author} {\bibfnamefont {Joao}\ \bibnamefont
  {Penedones}}, \bibinfo {author} {\bibfnamefont {Jonathan}\ \bibnamefont
  {Toledo}}, \bibinfo {author} {\bibfnamefont {Balt~C.}\ \bibnamefont {van
  Rees}}, \ and\ \bibinfo {author} {\bibfnamefont {Pedro}\ \bibnamefont
  {Vieira}},\ }\bibfield  {title} {\enquote {\bibinfo {title} {{The S-matrix
  bootstrap. Part III: higher dimensional amplitudes}},}\ }\href {\doibase
  10.1007/JHEP12(2019)040} {\bibfield  {journal} {\bibinfo  {journal} {JHEP}\
  }\textbf {\bibinfo {volume} {12}},\ \bibinfo {pages} {040} (\bibinfo {year}
  {2019})},\ \Eprint {http://arxiv.org/abs/1708.06765} {arXiv:1708.06765
  [hep-th]} \BibitemShut {NoStop}%
\bibitem [{\citenamefont {He}\ \emph {et~al.}(2018)\citenamefont {He},
  \citenamefont {Irrgang},\ and\ \citenamefont {Kruczenski}}]{He:2018uxa}%
  \BibitemOpen
  \bibfield  {author} {\bibinfo {author} {\bibfnamefont {Yifei}\ \bibnamefont
  {He}}, \bibinfo {author} {\bibfnamefont {Andrew}\ \bibnamefont {Irrgang}}, \
  and\ \bibinfo {author} {\bibfnamefont {Martin}\ \bibnamefont {Kruczenski}},\
  }\bibfield  {title} {\enquote {\bibinfo {title} {{A note on the S-matrix
  bootstrap for the 2d O(N) bosonic model}},}\ }\href {\doibase
  10.1007/JHEP11(2018)093} {\bibfield  {journal} {\bibinfo  {journal} {JHEP}\
  }\textbf {\bibinfo {volume} {11}},\ \bibinfo {pages} {093} (\bibinfo {year}
  {2018})},\ \Eprint {http://arxiv.org/abs/1805.02812} {arXiv:1805.02812
  [hep-th]} \BibitemShut {NoStop}%
\bibitem [{\citenamefont {C{\'o}rdova}\ and\ \citenamefont
  {Vieira}(2018)}]{Cordova:2018uop}%
  \BibitemOpen
  \bibfield  {author} {\bibinfo {author} {\bibfnamefont {Luc{\'\i}a}\
  \bibnamefont {C{\'o}rdova}}\ and\ \bibinfo {author} {\bibfnamefont {Pedro}\
  \bibnamefont {Vieira}},\ }\bibfield  {title} {\enquote {\bibinfo {title}
  {{Adding flavour to the S-matrix bootstrap}},}\ }\href {\doibase
  10.1007/JHEP12(2018)063} {\bibfield  {journal} {\bibinfo  {journal} {JHEP}\
  }\textbf {\bibinfo {volume} {12}},\ \bibinfo {pages} {063} (\bibinfo {year}
  {2018})},\ \Eprint {http://arxiv.org/abs/1805.11143} {arXiv:1805.11143
  [hep-th]} \BibitemShut {NoStop}%
\bibitem [{\citenamefont {Correia}\ \emph {et~al.}(2021)\citenamefont
  {Correia}, \citenamefont {Sever},\ and\ \citenamefont
  {Zhiboedov}}]{Correia:2020xtr}%
  \BibitemOpen
  \bibfield  {author} {\bibinfo {author} {\bibfnamefont {Miguel}\ \bibnamefont
  {Correia}}, \bibinfo {author} {\bibfnamefont {Amit}\ \bibnamefont {Sever}}, \
  and\ \bibinfo {author} {\bibfnamefont {Alexander}\ \bibnamefont
  {Zhiboedov}},\ }\bibfield  {title} {\enquote {\bibinfo {title} {{An
  analytical toolkit for the S-matrix bootstrap}},}\ }\href {\doibase
  10.1007/JHEP03(2021)013} {\bibfield  {journal} {\bibinfo  {journal} {JHEP}\
  }\textbf {\bibinfo {volume} {03}},\ \bibinfo {pages} {013} (\bibinfo {year}
  {2021})},\ \Eprint {http://arxiv.org/abs/2006.08221} {arXiv:2006.08221
  [hep-th]} \BibitemShut {NoStop}%
\bibitem [{\citenamefont {Kruczenski}\ \emph {et~al.}(2022)\citenamefont
  {Kruczenski}, \citenamefont {Penedones},\ and\ \citenamefont {van
  Rees}}]{Kruczenski:2022lot}%
  \BibitemOpen
  \bibfield  {author} {\bibinfo {author} {\bibfnamefont {Martin}\ \bibnamefont
  {Kruczenski}}, \bibinfo {author} {\bibfnamefont {Joao}\ \bibnamefont
  {Penedones}}, \ and\ \bibinfo {author} {\bibfnamefont {Balt~C.}\ \bibnamefont
  {van Rees}},\ }\bibfield  {title} {\enquote {\bibinfo {title} {{Snowmass
  White Paper: S-matrix Bootstrap}},}\ }\href@noop {} {\  (\bibinfo {year}
  {2022})},\ \Eprint {http://arxiv.org/abs/2203.02421} {arXiv:2203.02421
  [hep-th]} \BibitemShut {NoStop}%
\bibitem [{\citenamefont {Correia}\ \emph {et~al.}(2025)\citenamefont
  {Correia}, \citenamefont {Gopalka}, \citenamefont {Isabella},\ and\
  \citenamefont {Wolz}}]{Correia:2025enx}%
  \BibitemOpen
  \bibfield  {author} {\bibinfo {author} {\bibfnamefont {Miguel}\ \bibnamefont
  {Correia}}, \bibinfo {author} {\bibfnamefont {Tushar}\ \bibnamefont
  {Gopalka}}, \bibinfo {author} {\bibfnamefont {Giulia}\ \bibnamefont
  {Isabella}}, \ and\ \bibinfo {author} {\bibfnamefont {Anna~M.}\ \bibnamefont
  {Wolz}},\ }\bibfield  {title} {\enquote {\bibinfo {title} {{Analyticity of
  the Black Hole S-Matrix}},}\ }\href@noop {} {\  (\bibinfo {year} {2025})},\
  \Eprint {http://arxiv.org/abs/2511.11794} {arXiv:2511.11794 [hep-th]}
  \BibitemShut {NoStop}%
\bibitem [{\citenamefont {Gounaris}\ \emph {et~al.}(1995)\citenamefont
  {Gounaris}, \citenamefont {Layssac}, \citenamefont {Paschalis},\ and\
  \citenamefont {Renard}}]{Gounaris:1994cm}%
  \BibitemOpen
  \bibfield  {author} {\bibinfo {author} {\bibfnamefont {G.~J.}\ \bibnamefont
  {Gounaris}}, \bibinfo {author} {\bibfnamefont {J.}~\bibnamefont {Layssac}},
  \bibinfo {author} {\bibfnamefont {J.~E.}\ \bibnamefont {Paschalis}}, \ and\
  \bibinfo {author} {\bibfnamefont {F.~M.}\ \bibnamefont {Renard}},\ }\bibfield
   {title} {\enquote {\bibinfo {title} {{Unitarity constraints for new physics
  induced by dim-6 operators}},}\ }\href {\doibase 10.1007/BF01579637}
  {\bibfield  {journal} {\bibinfo  {journal} {Z. Phys. C}\ }\textbf {\bibinfo
  {volume} {66}},\ \bibinfo {pages} {619--632} (\bibinfo {year} {1995})},\
  \Eprint {http://arxiv.org/abs/hep-ph/9409260} {arXiv:hep-ph/9409260}
  \BibitemShut {NoStop}%
\bibitem [{\citenamefont {Corbett}\ \emph {et~al.}(2015)\citenamefont
  {Corbett}, \citenamefont {{\'E}boli},\ and\ \citenamefont
  {Gonzalez-Garcia}}]{Corbett:2014ora}%
  \BibitemOpen
  \bibfield  {author} {\bibinfo {author} {\bibfnamefont {Tyler}\ \bibnamefont
  {Corbett}}, \bibinfo {author} {\bibfnamefont {O.~J.~P.}\ \bibnamefont
  {{\'E}boli}}, \ and\ \bibinfo {author} {\bibfnamefont {M.~C.}\ \bibnamefont
  {Gonzalez-Garcia}},\ }\bibfield  {title} {\enquote {\bibinfo {title}
  {{Unitarity Constraints on Dimension-Six Operators}},}\ }\href {\doibase
  10.1103/PhysRevD.91.035014} {\bibfield  {journal} {\bibinfo  {journal} {Phys.
  Rev. D}\ }\textbf {\bibinfo {volume} {91}},\ \bibinfo {pages} {035014}
  (\bibinfo {year} {2015})},\ \Eprint {http://arxiv.org/abs/1411.5026}
  {arXiv:1411.5026 [hep-ph]} \BibitemShut {NoStop}%
\bibitem [{\citenamefont {Di~Luzio}\ \emph {et~al.}(2017)\citenamefont
  {Di~Luzio}, \citenamefont {Kamenik},\ and\ \citenamefont
  {Nardecchia}}]{DiLuzio:2016sur}%
  \BibitemOpen
  \bibfield  {author} {\bibinfo {author} {\bibfnamefont {Luca}\ \bibnamefont
  {Di~Luzio}}, \bibinfo {author} {\bibfnamefont {Jernej~F.}\ \bibnamefont
  {Kamenik}}, \ and\ \bibinfo {author} {\bibfnamefont {Marco}\ \bibnamefont
  {Nardecchia}},\ }\bibfield  {title} {\enquote {\bibinfo {title}
  {{Implications of perturbative unitarity for scalar di-boson resonance
  searches at LHC}},}\ }\href {\doibase 10.1140/epjc/s10052-017-4594-2}
  {\bibfield  {journal} {\bibinfo  {journal} {Eur. Phys. J. C}\ }\textbf
  {\bibinfo {volume} {77}},\ \bibinfo {pages} {30} (\bibinfo {year} {2017})},\
  \Eprint {http://arxiv.org/abs/1604.05746} {arXiv:1604.05746 [hep-ph]}
  \BibitemShut {NoStop}%
\bibitem [{\citenamefont {Corbett}\ \emph {et~al.}(2017)\citenamefont
  {Corbett}, \citenamefont {{\'E}boli},\ and\ \citenamefont
  {Gonzalez-Garcia}}]{Corbett:2017qgl}%
  \BibitemOpen
  \bibfield  {author} {\bibinfo {author} {\bibfnamefont {Tyler}\ \bibnamefont
  {Corbett}}, \bibinfo {author} {\bibfnamefont {O.~J.~P.}\ \bibnamefont
  {{\'E}boli}}, \ and\ \bibinfo {author} {\bibfnamefont {M.~C.}\ \bibnamefont
  {Gonzalez-Garcia}},\ }\bibfield  {title} {\enquote {\bibinfo {title}
  {{Unitarity Constraints on Dimension-six Operators II: Including Fermionic
  Operators}},}\ }\href {\doibase 10.1103/PhysRevD.96.035006} {\bibfield
  {journal} {\bibinfo  {journal} {Phys. Rev. D}\ }\textbf {\bibinfo {volume}
  {96}},\ \bibinfo {pages} {035006} (\bibinfo {year} {2017})},\ \Eprint
  {http://arxiv.org/abs/1705.09294} {arXiv:1705.09294 [hep-ph]} \BibitemShut
  {NoStop}%
\bibitem [{\citenamefont {Chang}\ and\ \citenamefont
  {Luty}(2020)}]{Chang:2019vez}%
  \BibitemOpen
  \bibfield  {author} {\bibinfo {author} {\bibfnamefont {Spencer}\ \bibnamefont
  {Chang}}\ and\ \bibinfo {author} {\bibfnamefont {Markus~A.}\ \bibnamefont
  {Luty}},\ }\bibfield  {title} {\enquote {\bibinfo {title} {{The Higgs
  Trilinear Coupling and the Scale of New Physics}},}\ }\href {\doibase
  10.1007/JHEP03(2020)140} {\bibfield  {journal} {\bibinfo  {journal} {JHEP}\
  }\textbf {\bibinfo {volume} {03}},\ \bibinfo {pages} {140} (\bibinfo {year}
  {2020})},\ \Eprint {http://arxiv.org/abs/1902.05556} {arXiv:1902.05556
  [hep-ph]} \BibitemShut {NoStop}%
\bibitem [{\citenamefont {Remmen}\ and\ \citenamefont
  {Rodd}(2019)}]{Remmen:2019cyz}%
  \BibitemOpen
  \bibfield  {author} {\bibinfo {author} {\bibfnamefont {Grant~N.}\
  \bibnamefont {Remmen}}\ and\ \bibinfo {author} {\bibfnamefont {Nicholas~L.}\
  \bibnamefont {Rodd}},\ }\bibfield  {title} {\enquote {\bibinfo {title}
  {{Consistency of the Standard Model Effective Field Theory}},}\ }\href
  {\doibase 10.1007/JHEP12(2019)032} {\bibfield  {journal} {\bibinfo  {journal}
  {JHEP}\ }\textbf {\bibinfo {volume} {12}},\ \bibinfo {pages} {032} (\bibinfo
  {year} {2019})},\ \Eprint {http://arxiv.org/abs/1908.09845} {arXiv:1908.09845
  [hep-ph]} \BibitemShut {NoStop}%
\bibitem [{\citenamefont {Remmen}\ and\ \citenamefont
  {Rodd}(2020)}]{Remmen:2020vts}%
  \BibitemOpen
  \bibfield  {author} {\bibinfo {author} {\bibfnamefont {Grant~N.}\
  \bibnamefont {Remmen}}\ and\ \bibinfo {author} {\bibfnamefont {Nicholas~L.}\
  \bibnamefont {Rodd}},\ }\bibfield  {title} {\enquote {\bibinfo {title}
  {{Flavor Constraints from Unitarity and Analyticity}},}\ }\href {\doibase
  10.1103/PhysRevLett.127.149901} {\bibfield  {journal} {\bibinfo  {journal}
  {Phys. Rev. Lett.}\ }\textbf {\bibinfo {volume} {125}},\ \bibinfo {pages}
  {081601} (\bibinfo {year} {2020})},\ \bibinfo {note} {[Erratum:
  Phys.Rev.Lett. 127, 149901 (2021)]},\ \Eprint
  {http://arxiv.org/abs/2004.02885} {arXiv:2004.02885 [hep-ph]} \BibitemShut
  {NoStop}%
\bibitem [{\citenamefont {Falkowski}\ and\ \citenamefont
  {Rattazzi}(2019)}]{Falkowski:2019tft}%
  \BibitemOpen
  \bibfield  {author} {\bibinfo {author} {\bibfnamefont {Adam}\ \bibnamefont
  {Falkowski}}\ and\ \bibinfo {author} {\bibfnamefont {Riccardo}\ \bibnamefont
  {Rattazzi}},\ }\bibfield  {title} {\enquote {\bibinfo {title} {{Which
  EFT}},}\ }\href {\doibase 10.1007/JHEP10(2019)255} {\bibfield  {journal}
  {\bibinfo  {journal} {JHEP}\ }\textbf {\bibinfo {volume} {10}},\ \bibinfo
  {pages} {255} (\bibinfo {year} {2019})},\ \Eprint
  {http://arxiv.org/abs/1902.05936} {arXiv:1902.05936 [hep-ph]} \BibitemShut
  {NoStop}%
\bibitem [{\citenamefont {Remmen}\ and\ \citenamefont
  {Rodd}(2022{\natexlab{a}})}]{Remmen:2020uze}%
  \BibitemOpen
  \bibfield  {author} {\bibinfo {author} {\bibfnamefont {Grant~N.}\
  \bibnamefont {Remmen}}\ and\ \bibinfo {author} {\bibfnamefont {Nicholas~L.}\
  \bibnamefont {Rodd}},\ }\bibfield  {title} {\enquote {\bibinfo {title}
  {{Signs, spin, SMEFT: Sum rules at dimension six}},}\ }\href {\doibase
  10.1103/PhysRevD.105.036006} {\bibfield  {journal} {\bibinfo  {journal}
  {Phys. Rev. D}\ }\textbf {\bibinfo {volume} {105}},\ \bibinfo {pages}
  {036006} (\bibinfo {year} {2022}{\natexlab{a}})},\ \Eprint
  {http://arxiv.org/abs/2010.04723} {arXiv:2010.04723 [hep-ph]} \BibitemShut
  {NoStop}%
\bibitem [{\citenamefont {Cohen}\ \emph {et~al.}(2022)\citenamefont {Cohen},
  \citenamefont {Doss},\ and\ \citenamefont {Lu}}]{Cohen:2021gdw}%
  \BibitemOpen
  \bibfield  {author} {\bibinfo {author} {\bibfnamefont {Timothy}\ \bibnamefont
  {Cohen}}, \bibinfo {author} {\bibfnamefont {Joel}\ \bibnamefont {Doss}}, \
  and\ \bibinfo {author} {\bibfnamefont {Xiaochuan}\ \bibnamefont {Lu}},\
  }\bibfield  {title} {\enquote {\bibinfo {title} {{Unitarity bounds on
  effective field theories at the LHC}},}\ }\href {\doibase
  10.1007/JHEP04(2022)155} {\bibfield  {journal} {\bibinfo  {journal} {JHEP}\
  }\textbf {\bibinfo {volume} {04}},\ \bibinfo {pages} {155} (\bibinfo {year}
  {2022})},\ \Eprint {http://arxiv.org/abs/2111.09895} {arXiv:2111.09895
  [hep-ph]} \BibitemShut {NoStop}%
\bibitem [{\citenamefont {Remmen}\ and\ \citenamefont
  {Rodd}(2022{\natexlab{b}})}]{Remmen:2022orj}%
  \BibitemOpen
  \bibfield  {author} {\bibinfo {author} {\bibfnamefont {Grant~N.}\
  \bibnamefont {Remmen}}\ and\ \bibinfo {author} {\bibfnamefont {Nicholas~L.}\
  \bibnamefont {Rodd}},\ }\bibfield  {title} {\enquote {\bibinfo {title}
  {{Spinning sum rules for the dimension-six SMEFT}},}\ }\href {\doibase
  10.1007/JHEP09(2022)030} {\bibfield  {journal} {\bibinfo  {journal} {JHEP}\
  }\textbf {\bibinfo {volume} {09}},\ \bibinfo {pages} {030} (\bibinfo {year}
  {2022}{\natexlab{b}})},\ \Eprint {http://arxiv.org/abs/2206.13524}
  {arXiv:2206.13524 [hep-ph]} \BibitemShut {NoStop}%
\bibitem [{\citenamefont {Cao}\ \emph {et~al.}(2025)\citenamefont {Cao},
  \citenamefont {Liu},\ and\ \citenamefont {Yuan}}]{Cao:2024vfc}%
  \BibitemOpen
  \bibfield  {author} {\bibinfo {author} {\bibfnamefont {Qing-Hong}\
  \bibnamefont {Cao}}, \bibinfo {author} {\bibfnamefont {Yandong}\ \bibnamefont
  {Liu}}, \ and\ \bibinfo {author} {\bibfnamefont {Shu-Run}\ \bibnamefont
  {Yuan}},\ }\bibfield  {title} {\enquote {\bibinfo {title} {{Unitarity bounds
  and basis transformations in SMEFT: An analysis of Warsaw and SILH bases}},}\
  }\href {\doibase 10.1016/j.nuclphysb.2024.116781} {\bibfield  {journal}
  {\bibinfo  {journal} {Nucl. Phys. B}\ }\textbf {\bibinfo {volume} {1010}},\
  \bibinfo {pages} {116781} (\bibinfo {year} {2025})}\BibitemShut {NoStop}%
\bibitem [{\citenamefont {Mahmud}\ and\ \citenamefont
  {Tobioka}(2024)}]{Mahmud:2024iyn}%
  \BibitemOpen
  \bibfield  {author} {\bibinfo {author} {\bibfnamefont {Shameran}\
  \bibnamefont {Mahmud}}\ and\ \bibinfo {author} {\bibfnamefont {Kohsaku}\
  \bibnamefont {Tobioka}},\ }\bibfield  {title} {\enquote {\bibinfo {title}
  {{Energy growth in V$_{L}$V$_{L}${\textrightarrow} V$_{L}$V$_{L}$,
  V$_{L}$V$_{L}$h scattering to probe Higgs cubic and HEFT interactions}},}\
  }\href {\doibase 10.1007/JHEP09(2024)073} {\bibfield  {journal} {\bibinfo
  {journal} {JHEP}\ }\textbf {\bibinfo {volume} {09}},\ \bibinfo {pages} {073}
  (\bibinfo {year} {2024})},\ \Eprint {http://arxiv.org/abs/2406.03522}
  {arXiv:2406.03522 [hep-ph]} \BibitemShut {NoStop}%
\bibitem [{\citenamefont {Remmen}\ and\ \citenamefont
  {Rodd}(2026)}]{Remmen:2024hry}%
  \BibitemOpen
  \bibfield  {author} {\bibinfo {author} {\bibfnamefont {Grant~N.}\
  \bibnamefont {Remmen}}\ and\ \bibinfo {author} {\bibfnamefont {Nicholas~L.}\
  \bibnamefont {Rodd}},\ }\bibfield  {title} {\enquote {\bibinfo {title}
  {{Positively identifying Higgs effective field theory or standard model
  effective field theory}},}\ }\href {\doibase 10.1103/vj7j-zj11} {\bibfield
  {journal} {\bibinfo  {journal} {Phys. Rev. D}\ }\textbf {\bibinfo {volume}
  {113}},\ \bibinfo {pages} {036027} (\bibinfo {year} {2026})},\ \Eprint
  {http://arxiv.org/abs/2412.07827} {arXiv:2412.07827 [hep-ph]} \BibitemShut
  {NoStop}%
\bibitem [{\citenamefont {Degrande}\ \emph {et~al.}(2025)\citenamefont
  {Degrande}, \citenamefont {Li},\ and\ \citenamefont {Xu}}]{Degrande:2025uil}%
  \BibitemOpen
  \bibfield  {author} {\bibinfo {author} {\bibfnamefont {C{\'e}line}\
  \bibnamefont {Degrande}}, \bibinfo {author} {\bibfnamefont {Hao-Lin}\
  \bibnamefont {Li}}, \ and\ \bibinfo {author} {\bibfnamefont {Ling-Xiao}\
  \bibnamefont {Xu}},\ }\bibfield  {title} {\enquote {\bibinfo {title}
  {{Partial-Wave Unitarity Bounds on Higher-Dimensional Operators from 2-to-$N$
  Scattering}},}\ }\href@noop {} {\  (\bibinfo {year} {2025})},\ \Eprint
  {http://arxiv.org/abs/2511.15524} {arXiv:2511.15524 [hep-ph]} \BibitemShut
  {NoStop}%
\bibitem [{\citenamefont {Bresciani}\ \emph {et~al.}(2025)\citenamefont
  {Bresciani}, \citenamefont {Levati},\ and\ \citenamefont
  {Paradisi}}]{Bresciani:2025toe}%
  \BibitemOpen
  \bibfield  {author} {\bibinfo {author} {\bibfnamefont {Luigi~C.}\
  \bibnamefont {Bresciani}}, \bibinfo {author} {\bibfnamefont {Gabriele}\
  \bibnamefont {Levati}}, \ and\ \bibinfo {author} {\bibfnamefont {Paride}\
  \bibnamefont {Paradisi}},\ }\bibfield  {title} {\enquote {\bibinfo {title}
  {{Amplitudes and partial wave unitarity bounds}},}\ }\href@noop {} {\
  (\bibinfo {year} {2025})},\ \Eprint {http://arxiv.org/abs/2504.12855}
  {arXiv:2504.12855 [hep-ph]} \BibitemShut {NoStop}%
\bibitem [{\citenamefont {Guerrieri}\ \emph {et~al.}(2021)\citenamefont
  {Guerrieri}, \citenamefont {Penedones},\ and\ \citenamefont
  {Vieira}}]{Guerrieri:2020bto}%
  \BibitemOpen
  \bibfield  {author} {\bibinfo {author} {\bibfnamefont {Andrea~L.}\
  \bibnamefont {Guerrieri}}, \bibinfo {author} {\bibfnamefont {Joao}\
  \bibnamefont {Penedones}}, \ and\ \bibinfo {author} {\bibfnamefont {Pedro}\
  \bibnamefont {Vieira}},\ }\bibfield  {title} {\enquote {\bibinfo {title}
  {{S-matrix bootstrap for effective field theories: massless pions}},}\ }\href
  {\doibase 10.1007/JHEP06(2021)088} {\bibfield  {journal} {\bibinfo  {journal}
  {JHEP}\ }\textbf {\bibinfo {volume} {06}},\ \bibinfo {pages} {088} (\bibinfo
  {year} {2021})},\ \Eprint {http://arxiv.org/abs/2011.02802} {arXiv:2011.02802
  [hep-th]} \BibitemShut {NoStop}%
\bibitem [{\citenamefont {Albert}\ and\ \citenamefont
  {Rastelli}(2022)}]{Albert:2022oes}%
  \BibitemOpen
  \bibfield  {author} {\bibinfo {author} {\bibfnamefont {Jan}\ \bibnamefont
  {Albert}}\ and\ \bibinfo {author} {\bibfnamefont {Leonardo}\ \bibnamefont
  {Rastelli}},\ }\bibfield  {title} {\enquote {\bibinfo {title} {{Bootstrapping
  pions at large N}},}\ }\href {\doibase 10.1007/JHEP08(2022)151} {\bibfield
  {journal} {\bibinfo  {journal} {JHEP}\ }\textbf {\bibinfo {volume} {08}},\
  \bibinfo {pages} {151} (\bibinfo {year} {2022})},\ \Eprint
  {http://arxiv.org/abs/2203.11950} {arXiv:2203.11950 [hep-th]} \BibitemShut
  {NoStop}%
\bibitem [{\citenamefont {Albert}\ and\ \citenamefont
  {Rastelli}(2024)}]{Albert:2023jtd}%
  \BibitemOpen
  \bibfield  {author} {\bibinfo {author} {\bibfnamefont {Jan}\ \bibnamefont
  {Albert}}\ and\ \bibinfo {author} {\bibfnamefont {Leonardo}\ \bibnamefont
  {Rastelli}},\ }\bibfield  {title} {\enquote {\bibinfo {title} {{Bootstrapping
  pions at large N. Part II. Background gauge fields and the chiral
  anomaly}},}\ }\href {\doibase 10.1007/JHEP09(2024)039} {\bibfield  {journal}
  {\bibinfo  {journal} {JHEP}\ }\textbf {\bibinfo {volume} {09}},\ \bibinfo
  {pages} {039} (\bibinfo {year} {2024})},\ \Eprint
  {http://arxiv.org/abs/2307.01246} {arXiv:2307.01246 [hep-th]} \BibitemShut
  {NoStop}%
\bibitem [{\citenamefont {Albert}\ \emph {et~al.}(2024)\citenamefont {Albert},
  \citenamefont {Henriksson}, \citenamefont {Rastelli},\ and\ \citenamefont
  {Vichi}}]{Albert:2023seb}%
  \BibitemOpen
  \bibfield  {author} {\bibinfo {author} {\bibfnamefont {Jan}\ \bibnamefont
  {Albert}}, \bibinfo {author} {\bibfnamefont {Johan}\ \bibnamefont
  {Henriksson}}, \bibinfo {author} {\bibfnamefont {Leonardo}\ \bibnamefont
  {Rastelli}}, \ and\ \bibinfo {author} {\bibfnamefont {Alessandro}\
  \bibnamefont {Vichi}},\ }\bibfield  {title} {\enquote {\bibinfo {title}
  {{Bootstrapping mesons at large N: Regge trajectory from spin-two
  maximization}},}\ }\href {\doibase 10.1007/JHEP09(2024)172} {\bibfield
  {journal} {\bibinfo  {journal} {JHEP}\ }\textbf {\bibinfo {volume} {09}},\
  \bibinfo {pages} {172} (\bibinfo {year} {2024})},\ \Eprint
  {http://arxiv.org/abs/2312.15013} {arXiv:2312.15013 [hep-th]} \BibitemShut
  {NoStop}%
\bibitem [{\citenamefont {Yennie}\ \emph {et~al.}(1961)\citenamefont {Yennie},
  \citenamefont {Frautschi},\ and\ \citenamefont {Suura}}]{Yennie:1961ad}%
  \BibitemOpen
  \bibfield  {author} {\bibinfo {author} {\bibfnamefont {D.~R.}\ \bibnamefont
  {Yennie}}, \bibinfo {author} {\bibfnamefont {Steven~C.}\ \bibnamefont
  {Frautschi}}, \ and\ \bibinfo {author} {\bibfnamefont {H.}~\bibnamefont
  {Suura}},\ }\bibfield  {title} {\enquote {\bibinfo {title} {{The infrared
  divergence phenomena and high-energy processes}},}\ }\href {\doibase
  10.1016/0003-4916(61)90151-8} {\bibfield  {journal} {\bibinfo  {journal}
  {Annals Phys.}\ }\textbf {\bibinfo {volume} {13}},\ \bibinfo {pages}
  {379--452} (\bibinfo {year} {1961})}\BibitemShut {NoStop}%
\bibitem [{\citenamefont {Weinberg}(1965)}]{Weinberg:1965nx}%
  \BibitemOpen
  \bibfield  {author} {\bibinfo {author} {\bibfnamefont {Steven}\ \bibnamefont
  {Weinberg}},\ }\bibfield  {title} {\enquote {\bibinfo {title} {{Infrared
  photons and gravitons}},}\ }\href {\doibase 10.1103/PhysRev.140.B516}
  {\bibfield  {journal} {\bibinfo  {journal} {Phys. Rev.}\ }\textbf {\bibinfo
  {volume} {140}},\ \bibinfo {pages} {B516--B524} (\bibinfo {year}
  {1965})}\BibitemShut {NoStop}%
\bibitem [{\citenamefont {Libby}\ and\ \citenamefont
  {Sterman}(1978)}]{Libby:1978qf}%
  \BibitemOpen
  \bibfield  {author} {\bibinfo {author} {\bibfnamefont {Stephen~B.}\
  \bibnamefont {Libby}}\ and\ \bibinfo {author} {\bibfnamefont {George~F.}\
  \bibnamefont {Sterman}},\ }\bibfield  {title} {\enquote {\bibinfo {title}
  {{Jet and Lepton Pair Production in High-Energy Lepton-Hadron and
  Hadron-Hadron Scattering}},}\ }\href {\doibase 10.1103/PhysRevD.18.3252}
  {\bibfield  {journal} {\bibinfo  {journal} {Phys. Rev. D}\ }\textbf {\bibinfo
  {volume} {18}},\ \bibinfo {pages} {3252} (\bibinfo {year}
  {1978})}\BibitemShut {NoStop}%
\bibitem [{\citenamefont {Becher}\ and\ \citenamefont
  {Neubert}(2009)}]{Becher:2009qa}%
  \BibitemOpen
  \bibfield  {author} {\bibinfo {author} {\bibfnamefont {Thomas}\ \bibnamefont
  {Becher}}\ and\ \bibinfo {author} {\bibfnamefont {Matthias}\ \bibnamefont
  {Neubert}},\ }\bibfield  {title} {\enquote {\bibinfo {title} {{On the
  Structure of Infrared Singularities of Gauge-Theory Amplitudes}},}\ }\href
  {\doibase 10.1088/1126-6708/2009/06/081} {\bibfield  {journal} {\bibinfo
  {journal} {JHEP}\ }\textbf {\bibinfo {volume} {06}},\ \bibinfo {pages} {081}
  (\bibinfo {year} {2009})},\ \bibinfo {note} {[Erratum: JHEP 11, 024
  (2013)]},\ \Eprint {http://arxiv.org/abs/0903.1126} {arXiv:0903.1126
  [hep-ph]} \BibitemShut {NoStop}%
\bibitem [{\citenamefont {Campiglia}\ and\ \citenamefont
  {Laddha}(2015)}]{Campiglia:2015qka}%
  \BibitemOpen
  \bibfield  {author} {\bibinfo {author} {\bibfnamefont {Miguel}\ \bibnamefont
  {Campiglia}}\ and\ \bibinfo {author} {\bibfnamefont {Alok}\ \bibnamefont
  {Laddha}},\ }\bibfield  {title} {\enquote {\bibinfo {title} {{Asymptotic
  symmetries of QED and Weinberg{\textquoteright}s soft photon theorem}},}\
  }\href {\doibase 10.1007/JHEP07(2015)115} {\bibfield  {journal} {\bibinfo
  {journal} {JHEP}\ }\textbf {\bibinfo {volume} {07}},\ \bibinfo {pages} {115}
  (\bibinfo {year} {2015})},\ \Eprint {http://arxiv.org/abs/1505.05346}
  {arXiv:1505.05346 [hep-th]} \BibitemShut {NoStop}%
\bibitem [{\citenamefont {Kapec}\ \emph {et~al.}(2017)\citenamefont {Kapec},
  \citenamefont {Pate},\ and\ \citenamefont {Strominger}}]{Kapec:2015ena}%
  \BibitemOpen
  \bibfield  {author} {\bibinfo {author} {\bibfnamefont {Daniel}\ \bibnamefont
  {Kapec}}, \bibinfo {author} {\bibfnamefont {Monica}\ \bibnamefont {Pate}}, \
  and\ \bibinfo {author} {\bibfnamefont {Andrew}\ \bibnamefont {Strominger}},\
  }\bibfield  {title} {\enquote {\bibinfo {title} {{New Symmetries of QED}},}\
  }\href {\doibase 10.4310/ATMP.2017.v21.n7.a7} {\bibfield  {journal} {\bibinfo
   {journal} {Adv. Theor. Math. Phys.}\ }\textbf {\bibinfo {volume} {21}},\
  \bibinfo {pages} {1769--1785} (\bibinfo {year} {2017})},\ \Eprint
  {http://arxiv.org/abs/1506.02906} {arXiv:1506.02906 [hep-th]} \BibitemShut
  {NoStop}%
\bibitem [{\citenamefont {Bellazzini}\ \emph {et~al.}(2022)\citenamefont
  {Bellazzini}, \citenamefont {Riembau},\ and\ \citenamefont
  {Riva}}]{Bellazzini:2021oaj}%
  \BibitemOpen
  \bibfield  {author} {\bibinfo {author} {\bibfnamefont {Brando}\ \bibnamefont
  {Bellazzini}}, \bibinfo {author} {\bibfnamefont {Marc}\ \bibnamefont
  {Riembau}}, \ and\ \bibinfo {author} {\bibfnamefont {Francesco}\ \bibnamefont
  {Riva}},\ }\bibfield  {title} {\enquote {\bibinfo {title} {{IR side of
  positivity bounds}},}\ }\href {\doibase 10.1103/PhysRevD.106.105008}
  {\bibfield  {journal} {\bibinfo  {journal} {Phys. Rev. D}\ }\textbf {\bibinfo
  {volume} {106}},\ \bibinfo {pages} {105008} (\bibinfo {year} {2022})},\
  \Eprint {http://arxiv.org/abs/2112.12561} {arXiv:2112.12561 [hep-th]}
  \BibitemShut {NoStop}%
\bibitem [{\citenamefont {Caron-Huot}\ \emph
  {et~al.}(2023{\natexlab{a}})\citenamefont {Caron-Huot}, \citenamefont {Li},
  \citenamefont {Parra-Martinez},\ and\ \citenamefont
  {Simmons-Duffin}}]{Caron-Huot:2022ugt}%
  \BibitemOpen
  \bibfield  {author} {\bibinfo {author} {\bibfnamefont {Simon}\ \bibnamefont
  {Caron-Huot}}, \bibinfo {author} {\bibfnamefont {Yue-Zhou}\ \bibnamefont
  {Li}}, \bibinfo {author} {\bibfnamefont {Julio}\ \bibnamefont
  {Parra-Martinez}}, \ and\ \bibinfo {author} {\bibfnamefont {David}\
  \bibnamefont {Simmons-Duffin}},\ }\bibfield  {title} {\enquote {\bibinfo
  {title} {{Causality constraints on corrections to Einstein gravity}},}\
  }\href {\doibase 10.1007/JHEP05(2023)122} {\bibfield  {journal} {\bibinfo
  {journal} {JHEP}\ }\textbf {\bibinfo {volume} {05}},\ \bibinfo {pages} {122}
  (\bibinfo {year} {2023}{\natexlab{a}})},\ \Eprint
  {http://arxiv.org/abs/2201.06602} {arXiv:2201.06602 [hep-th]} \BibitemShut
  {NoStop}%
\bibitem [{\citenamefont {Caron-Huot}\ \emph
  {et~al.}(2023{\natexlab{b}})\citenamefont {Caron-Huot}, \citenamefont {Li},
  \citenamefont {Parra-Martinez},\ and\ \citenamefont
  {Simmons-Duffin}}]{Caron-Huot:2022jli}%
  \BibitemOpen
  \bibfield  {author} {\bibinfo {author} {\bibfnamefont {Simon}\ \bibnamefont
  {Caron-Huot}}, \bibinfo {author} {\bibfnamefont {Yue-Zhou}\ \bibnamefont
  {Li}}, \bibinfo {author} {\bibfnamefont {Julio}\ \bibnamefont
  {Parra-Martinez}}, \ and\ \bibinfo {author} {\bibfnamefont {David}\
  \bibnamefont {Simmons-Duffin}},\ }\bibfield  {title} {\enquote {\bibinfo
  {title} {{Graviton partial waves and causality in higher dimensions}},}\
  }\href {\doibase 10.1103/PhysRevD.108.026007} {\bibfield  {journal} {\bibinfo
   {journal} {Phys. Rev. D}\ }\textbf {\bibinfo {volume} {108}},\ \bibinfo
  {pages} {026007} (\bibinfo {year} {2023}{\natexlab{b}})},\ \Eprint
  {http://arxiv.org/abs/2205.01495} {arXiv:2205.01495 [hep-th]} \BibitemShut
  {NoStop}%
\bibitem [{\citenamefont {Chang}\ and\ \citenamefont
  {Parra-Martinez}(2025)}]{Chang:2025cxc}%
  \BibitemOpen
  \bibfield  {author} {\bibinfo {author} {\bibfnamefont {Cyuan-Han}\
  \bibnamefont {Chang}}\ and\ \bibinfo {author} {\bibfnamefont {Julio}\
  \bibnamefont {Parra-Martinez}},\ }\bibfield  {title} {\enquote {\bibinfo
  {title} {{Graviton loops and negativity}},}\ }\href {\doibase
  10.1007/JHEP08(2025)175} {\bibfield  {journal} {\bibinfo  {journal} {JHEP}\
  }\textbf {\bibinfo {volume} {08}},\ \bibinfo {pages} {175} (\bibinfo {year}
  {2025})},\ \Eprint {http://arxiv.org/abs/2501.17949} {arXiv:2501.17949
  [hep-th]} \BibitemShut {NoStop}%
\bibitem [{\citenamefont {Bellazzini}\ \emph {et~al.}(2025)\citenamefont
  {Bellazzini}, \citenamefont {Berman}, \citenamefont {Isabella}, \citenamefont
  {Riva}, \citenamefont {Romano},\ and\ \citenamefont
  {Sciotti}}]{Bellazzini:2025bay}%
  \BibitemOpen
  \bibfield  {author} {\bibinfo {author} {\bibfnamefont {B.}~\bibnamefont
  {Bellazzini}}, \bibinfo {author} {\bibfnamefont {J.}~\bibnamefont {Berman}},
  \bibinfo {author} {\bibfnamefont {G.}~\bibnamefont {Isabella}}, \bibinfo
  {author} {\bibfnamefont {F.}~\bibnamefont {Riva}}, \bibinfo {author}
  {\bibfnamefont {M.}~\bibnamefont {Romano}}, \ and\ \bibinfo {author}
  {\bibfnamefont {F.}~\bibnamefont {Sciotti}},\ }\bibfield  {title} {\enquote
  {\bibinfo {title} {{Positivity with Long-Range Interactions}},}\ }\href@noop
  {} {\  (\bibinfo {year} {2025})},\ \Eprint {http://arxiv.org/abs/2512.13780}
  {arXiv:2512.13780 [hep-th]} \BibitemShut {NoStop}%
\bibitem [{\citenamefont {Jacob}\ and\ \citenamefont
  {Wick}(1959)}]{Jacob:1959at}%
  \BibitemOpen
  \bibfield  {author} {\bibinfo {author} {\bibfnamefont {M.}~\bibnamefont
  {Jacob}}\ and\ \bibinfo {author} {\bibfnamefont {G.~C.}\ \bibnamefont
  {Wick}},\ }\bibfield  {title} {\enquote {\bibinfo {title} {{On the General
  Theory of Collisions for Particles with Spin}},}\ }\href {\doibase
  10.1006/aphy.2000.6022} {\bibfield  {journal} {\bibinfo  {journal} {Annals
  Phys.}\ }\textbf {\bibinfo {volume} {7}},\ \bibinfo {pages} {404--428}
  (\bibinfo {year} {1959})}\BibitemShut {NoStop}%
\bibitem [{\citenamefont {Itzykson}\ and\ \citenamefont
  {Zuber}(1980)}]{Itzykson:1980rh}%
  \BibitemOpen
  \bibfield  {author} {\bibinfo {author} {\bibfnamefont {C.}~\bibnamefont
  {Itzykson}}\ and\ \bibinfo {author} {\bibfnamefont {J.~B.}\ \bibnamefont
  {Zuber}},\ }\href@noop {} {\emph {\bibinfo {title} {{Quantum Field
  Theory}}}},\ International Series In Pure and Applied Physics\ (\bibinfo
  {publisher} {McGraw-Hill},\ \bibinfo {address} {New York},\ \bibinfo {year}
  {1980})\BibitemShut {NoStop}%
\bibitem [{\citenamefont {Dollard}(1964)}]{Dollard:1964cok}%
  \BibitemOpen
  \bibfield  {author} {\bibinfo {author} {\bibfnamefont {John~D.}\ \bibnamefont
  {Dollard}},\ }\bibfield  {title} {\enquote {\bibinfo {title} {{Asymptotic
  Convergence and the Coulomb Interaction}},}\ }\href {\doibase
  10.1063/1.1704171} {\bibfield  {journal} {\bibinfo  {journal} {J. Math.
  Phys.}\ }\textbf {\bibinfo {volume} {5}},\ \bibinfo {pages} {729} (\bibinfo
  {year} {1964})}\BibitemShut {NoStop}%
\bibitem [{\citenamefont {Chung}(1965)}]{Chung:1965zza}%
  \BibitemOpen
  \bibfield  {author} {\bibinfo {author} {\bibfnamefont {Victor}\ \bibnamefont
  {Chung}},\ }\bibfield  {title} {\enquote {\bibinfo {title} {{Infrared
  Divergence in Quantum Electrodynamics}},}\ }\href {\doibase
  10.1103/PhysRev.140.B1110} {\bibfield  {journal} {\bibinfo  {journal} {Phys.
  Rev.}\ }\textbf {\bibinfo {volume} {140}},\ \bibinfo {pages} {B1110--B1122}
  (\bibinfo {year} {1965})}\BibitemShut {NoStop}%
\bibitem [{\citenamefont {Kibble}(1968)}]{Kibble:1968sfb}%
  \BibitemOpen
  \bibfield  {author} {\bibinfo {author} {\bibfnamefont {T.~W.~B.}\
  \bibnamefont {Kibble}},\ }\bibfield  {title} {\enquote {\bibinfo {title}
  {{Coherent Soft-Photon States and Infrared Divergences. I. Classical
  Currents}},}\ }\href {\doibase 10.1063/1.1664582} {\bibfield  {journal}
  {\bibinfo  {journal} {J. Math. Phys.}\ }\textbf {\bibinfo {volume} {9}},\
  \bibinfo {pages} {315--324} (\bibinfo {year} {1968})}\BibitemShut {NoStop}%
\bibitem [{\citenamefont {Kulish}\ and\ \citenamefont
  {Faddeev}(1970)}]{Kulish:1970ut}%
  \BibitemOpen
  \bibfield  {author} {\bibinfo {author} {\bibfnamefont {P.~P.}\ \bibnamefont
  {Kulish}}\ and\ \bibinfo {author} {\bibfnamefont {L.~D.}\ \bibnamefont
  {Faddeev}},\ }\bibfield  {title} {\enquote {\bibinfo {title} {{Asymptotic
  conditions and infrared divergences in quantum electrodynamics}},}\ }\href
  {\doibase 10.1007/BF01066485} {\bibfield  {journal} {\bibinfo  {journal}
  {Theor. Math. Phys.}\ }\textbf {\bibinfo {volume} {4}},\ \bibinfo {pages}
  {745} (\bibinfo {year} {1970})}\BibitemShut {NoStop}%
\bibitem [{\citenamefont {Dollard}(1971)}]{Dollard:1971qm}%
  \BibitemOpen
  \bibfield  {author} {\bibinfo {author} {\bibfnamefont {John~D.}\ \bibnamefont
  {Dollard}},\ }\bibfield  {title} {\enquote {\bibinfo {title}
  {{Quantum-Mechanical Scattering Theory for Short-Range and Coulomb
  Interactions}},}\ }\href {https://www.jstor.org/stable/44236056} {\bibfield
  {journal} {\bibinfo  {journal} {Rocky Mt. J. Math.}\ }\textbf {\bibinfo
  {volume} {1}},\ \bibinfo {pages} {5--88} (\bibinfo {year}
  {1971})}\BibitemShut {NoStop}%
\bibitem [{\citenamefont {Hannesdottir}\ and\ \citenamefont
  {Schwartz}(2020)}]{Hannesdottir:2019opa}%
  \BibitemOpen
  \bibfield  {author} {\bibinfo {author} {\bibfnamefont {Holmfridur}\
  \bibnamefont {Hannesdottir}}\ and\ \bibinfo {author} {\bibfnamefont
  {Matthew~D.}\ \bibnamefont {Schwartz}},\ }\bibfield  {title} {\enquote
  {\bibinfo {title} {{$S$ -Matrix for massless particles}},}\ }\href {\doibase
  10.1103/PhysRevD.101.105001} {\bibfield  {journal} {\bibinfo  {journal}
  {Phys. Rev. D}\ }\textbf {\bibinfo {volume} {101}},\ \bibinfo {pages}
  {105001} (\bibinfo {year} {2020})},\ \Eprint
  {http://arxiv.org/abs/1911.06821} {arXiv:1911.06821 [hep-th]} \BibitemShut
  {NoStop}%
\bibitem [{\citenamefont {Hannesdottir}\ and\ \citenamefont
  {Schwartz}(2023)}]{Hannesdottir:2019umk}%
  \BibitemOpen
  \bibfield  {author} {\bibinfo {author} {\bibfnamefont {Holmfridur}\
  \bibnamefont {Hannesdottir}}\ and\ \bibinfo {author} {\bibfnamefont
  {Matthew~D.}\ \bibnamefont {Schwartz}},\ }\bibfield  {title} {\enquote
  {\bibinfo {title} {{Finite $S$ matrix}},}\ }\href {\doibase
  10.1103/PhysRevD.107.L021701} {\bibfield  {journal} {\bibinfo  {journal}
  {Phys. Rev. D}\ }\textbf {\bibinfo {volume} {107}},\ \bibinfo {pages}
  {L021701} (\bibinfo {year} {2023})},\ \Eprint
  {http://arxiv.org/abs/1906.03271} {arXiv:1906.03271 [hep-th]} \BibitemShut
  {NoStop}%
\bibitem [{\citenamefont {Carney}\ \emph {et~al.}(2017)\citenamefont {Carney},
  \citenamefont {Chaurette}, \citenamefont {Neuenfeld},\ and\ \citenamefont
  {Semenoff}}]{Carney:2017jut}%
  \BibitemOpen
  \bibfield  {author} {\bibinfo {author} {\bibfnamefont {Daniel}\ \bibnamefont
  {Carney}}, \bibinfo {author} {\bibfnamefont {Laurent}\ \bibnamefont
  {Chaurette}}, \bibinfo {author} {\bibfnamefont {Dominik}\ \bibnamefont
  {Neuenfeld}}, \ and\ \bibinfo {author} {\bibfnamefont {Gordon~Walter}\
  \bibnamefont {Semenoff}},\ }\bibfield  {title} {\enquote {\bibinfo {title}
  {{Infrared quantum information}},}\ }\href {\doibase
  10.1103/PhysRevLett.119.180502} {\bibfield  {journal} {\bibinfo  {journal}
  {Phys. Rev. Lett.}\ }\textbf {\bibinfo {volume} {119}},\ \bibinfo {pages}
  {180502} (\bibinfo {year} {2017})},\ \Eprint
  {http://arxiv.org/abs/1706.03782} {arXiv:1706.03782 [hep-th]} \BibitemShut
  {NoStop}%
\bibitem [{\citenamefont {Carney}\ \emph
  {et~al.}(2018{\natexlab{a}})\citenamefont {Carney}, \citenamefont
  {Chaurette}, \citenamefont {Neuenfeld},\ and\ \citenamefont
  {Semenoff}}]{Carney:2017oxp}%
  \BibitemOpen
  \bibfield  {author} {\bibinfo {author} {\bibfnamefont {Daniel}\ \bibnamefont
  {Carney}}, \bibinfo {author} {\bibfnamefont {Laurent}\ \bibnamefont
  {Chaurette}}, \bibinfo {author} {\bibfnamefont {Dominik}\ \bibnamefont
  {Neuenfeld}}, \ and\ \bibinfo {author} {\bibfnamefont {Gordon~Walter}\
  \bibnamefont {Semenoff}},\ }\bibfield  {title} {\enquote {\bibinfo {title}
  {{Dressed infrared quantum information}},}\ }\href {\doibase
  10.1103/PhysRevD.97.025007} {\bibfield  {journal} {\bibinfo  {journal} {Phys.
  Rev. D}\ }\textbf {\bibinfo {volume} {97}},\ \bibinfo {pages} {025007}
  (\bibinfo {year} {2018}{\natexlab{a}})},\ \Eprint
  {http://arxiv.org/abs/1710.02531} {arXiv:1710.02531 [hep-th]} \BibitemShut
  {NoStop}%
\bibitem [{\citenamefont {Carney}\ \emph
  {et~al.}(2018{\natexlab{b}})\citenamefont {Carney}, \citenamefont
  {Chaurette}, \citenamefont {Neuenfeld},\ and\ \citenamefont
  {Semenoff}}]{Carney:2018ygh}%
  \BibitemOpen
  \bibfield  {author} {\bibinfo {author} {\bibfnamefont {Daniel}\ \bibnamefont
  {Carney}}, \bibinfo {author} {\bibfnamefont {Laurent}\ \bibnamefont
  {Chaurette}}, \bibinfo {author} {\bibfnamefont {Dominik}\ \bibnamefont
  {Neuenfeld}}, \ and\ \bibinfo {author} {\bibfnamefont {Gordon}\ \bibnamefont
  {Semenoff}},\ }\bibfield  {title} {\enquote {\bibinfo {title} {{On the need
  for soft dressing}},}\ }\href {\doibase 10.1007/JHEP09(2018)121} {\bibfield
  {journal} {\bibinfo  {journal} {JHEP}\ }\textbf {\bibinfo {volume} {09}},\
  \bibinfo {pages} {121} (\bibinfo {year} {2018}{\natexlab{b}})},\ \Eprint
  {http://arxiv.org/abs/1803.02370} {arXiv:1803.02370 [hep-th]} \BibitemShut
  {NoStop}%
\bibitem [{\citenamefont {Lippstreu}(2023)}]{Lippstreu:2023vvg}%
  \BibitemOpen
  \bibfield  {author} {\bibinfo {author} {\bibfnamefont {Luke}\ \bibnamefont
  {Lippstreu}},\ }\bibfield  {title} {\enquote {\bibinfo {title} {{A
  perturbation theory for the Coulomb phase infrared-divergence}},}\
  }\href@noop {} {\  (\bibinfo {year} {2023})},\ \Eprint
  {http://arxiv.org/abs/2312.08455} {arXiv:2312.08455 [hep-th]} \BibitemShut
  {NoStop}%
\bibitem [{\citenamefont {Lippstreu}(2025)}]{Lippstreu:2025jit}%
  \BibitemOpen
  \bibfield  {author} {\bibinfo {author} {\bibfnamefont {Luke}\ \bibnamefont
  {Lippstreu}},\ }\bibfield  {title} {\enquote {\bibinfo {title} {{Analytic
  Properties of Infrared-Finite Amplitudes in Theories with Long-Range
  Forces}},}\ }\href@noop {} {\  (\bibinfo {year} {2025})},\ \Eprint
  {http://arxiv.org/abs/2505.04702} {arXiv:2505.04702 [hep-th]} \BibitemShut
  {NoStop}%
\bibitem [{\citenamefont {Landau}\ and\ \citenamefont
  {Lifshits}(1991)}]{Landau:1991wop}%
  \BibitemOpen
  \bibfield  {author} {\bibinfo {author} {\bibfnamefont {Lev~Davidovich}\
  \bibnamefont {Landau}}\ and\ \bibinfo {author} {\bibfnamefont {E.~M.}\
  \bibnamefont {Lifshits}},\ }\href {\doibase 10.1016/C2013-0-02793-4} {\emph
  {\bibinfo {title} {{Quantum Mechanics}: {Non-Relativistic Theory}}}},\
  \bibinfo {series} {Course of Theoretical Physics}, Vol.\ \bibinfo {volume}
  {v.3}\ (\bibinfo  {publisher} {Butterworth-Heinemann},\ \bibinfo {address}
  {Oxford},\ \bibinfo {year} {1991})\BibitemShut {NoStop}%
\bibitem [{\citenamefont {Fuentes~Zamoro}\ \emph {et~al.}(2025)\citenamefont
  {Fuentes~Zamoro}, \citenamefont {Grinstein},\ and\ \citenamefont
  {Qu{\'\i}lez}}]{FuentesZamoro:2025exp}%
  \BibitemOpen
  \bibfield  {author} {\bibinfo {author} {\bibfnamefont {Marta}\ \bibnamefont
  {Fuentes~Zamoro}}, \bibinfo {author} {\bibfnamefont {Benjam{\'\i}n}\
  \bibnamefont {Grinstein}}, \ and\ \bibinfo {author} {\bibfnamefont {Pablo}\
  \bibnamefont {Qu{\'\i}lez}},\ }\bibfield  {title} {\enquote {\bibinfo {title}
  {{Taming forward scattering singularities in partial waves}},}\ }\href@noop
  {} {\  (\bibinfo {year} {2025})},\ \Eprint {http://arxiv.org/abs/2510.08784}
  {arXiv:2510.08784 [hep-ph]} \BibitemShut {NoStop}%
\bibitem [{\citenamefont {Taylor}(2012)}]{taylor2012scattering}%
  \BibitemOpen
  \bibfield  {author} {\bibinfo {author} {\bibfnamefont {J.R.}\ \bibnamefont
  {Taylor}},\ }\href {https://books.google.ch/books?id=OIaXvuwZMLQC} {\emph
  {\bibinfo {title} {Scattering Theory: The Quantum Theory of Nonrelativistic
  Collisions}}},\ Dover Books on Engineering\ (\bibinfo  {publisher} {Dover
  Publications},\ \bibinfo {year} {2012})\BibitemShut {NoStop}%
\bibitem [{\citenamefont {Weinberg}(2005)}]{Weinberg:1995mt}%
  \BibitemOpen
  \bibfield  {author} {\bibinfo {author} {\bibfnamefont {Steven}\ \bibnamefont
  {Weinberg}},\ }\href {\doibase 10.1017/CBO9781139644167} {\emph {\bibinfo
  {title} {{The Quantum theory of fields. Vol. 1: Foundations}}}}\ (\bibinfo
  {publisher} {Cambridge University Press},\ \bibinfo {year}
  {2005})\BibitemShut {NoStop}%
\bibitem [{\citenamefont {Beneke}\ and\ \citenamefont
  {Smirnov}(1998)}]{Beneke:1997zp}%
  \BibitemOpen
  \bibfield  {author} {\bibinfo {author} {\bibfnamefont {M.}~\bibnamefont
  {Beneke}}\ and\ \bibinfo {author} {\bibfnamefont {Vladimir~A.}\ \bibnamefont
  {Smirnov}},\ }\bibfield  {title} {\enquote {\bibinfo {title} {{Asymptotic
  expansion of Feynman integrals near threshold}},}\ }\href {\doibase
  10.1016/S0550-3213(98)00138-2} {\bibfield  {journal} {\bibinfo  {journal}
  {Nucl. Phys. B}\ }\textbf {\bibinfo {volume} {522}},\ \bibinfo {pages}
  {321--344} (\bibinfo {year} {1998})},\ \Eprint
  {http://arxiv.org/abs/hep-ph/9711391} {arXiv:hep-ph/9711391} \BibitemShut
  {NoStop}%
\bibitem [{\citenamefont {Messiah}(2014)}]{messiah2014quantum}%
  \BibitemOpen
  \bibfield  {author} {\bibinfo {author} {\bibfnamefont {A.}~\bibnamefont
  {Messiah}},\ }\href {https://books.google.com/books?id=voUUAwAAQBAJ} {\emph
  {\bibinfo {title} {Quantum Mechanics}}},\ Dover Books on Physics\ (\bibinfo
  {publisher} {Dover Publications},\ \bibinfo {year} {2014})\BibitemShut
  {NoStop}%
\bibitem [{\citenamefont {Rose}(1961)}]{rose1961relativistic}%
  \BibitemOpen
  \bibfield  {author} {\bibinfo {author} {\bibfnamefont {M.E.}\ \bibnamefont
  {Rose}},\ }\href {https://books.google.com/books?id=gh5RAAAAMAAJ} {\emph
  {\bibinfo {title} {Relativistic Electron Theory}}}\ (\bibinfo  {publisher}
  {Wiley},\ \bibinfo {year} {1961})\BibitemShut {NoStop}%
\bibitem [{\citenamefont {Rothstein}\ and\ \citenamefont
  {Stewart}(2016)}]{Rothstein:2016bsq}%
  \BibitemOpen
  \bibfield  {author} {\bibinfo {author} {\bibfnamefont {Ira~Z.}\ \bibnamefont
  {Rothstein}}\ and\ \bibinfo {author} {\bibfnamefont {Iain~W.}\ \bibnamefont
  {Stewart}},\ }\bibfield  {title} {\enquote {\bibinfo {title} {{An Effective
  Field Theory for Forward Scattering and Factorization Violation}},}\ }\href
  {\doibase 10.1007/JHEP08(2016)025} {\bibfield  {journal} {\bibinfo  {journal}
  {JHEP}\ }\textbf {\bibinfo {volume} {08}},\ \bibinfo {pages} {025} (\bibinfo
  {year} {2016})},\ \Eprint {http://arxiv.org/abs/1601.04695} {arXiv:1601.04695
  [hep-ph]} \BibitemShut {NoStop}%
\bibitem [{\citenamefont {Rothstein}\ and\ \citenamefont
  {Saavedra}(2024)}]{Rothstein:2024nlq}%
  \BibitemOpen
  \bibfield  {author} {\bibinfo {author} {\bibfnamefont {Ira~Z.}\ \bibnamefont
  {Rothstein}}\ and\ \bibinfo {author} {\bibfnamefont {Michael}\ \bibnamefont
  {Saavedra}},\ }\bibfield  {title} {\enquote {\bibinfo {title} {{A Systematic
  Lagrangian Formulation for Quantum and Classical Gravity at High
  Energies}},}\ }\href@noop {} {\  (\bibinfo {year} {2024})},\ \Eprint
  {http://arxiv.org/abs/2412.04428} {arXiv:2412.04428 [hep-th]} \BibitemShut
  {NoStop}%
\bibitem [{\citenamefont {Manohar}\ and\ \citenamefont
  {Wise}(2000)}]{Manohar:2000dt}%
  \BibitemOpen
  \bibfield  {author} {\bibinfo {author} {\bibfnamefont {Aneesh~V.}\
  \bibnamefont {Manohar}}\ and\ \bibinfo {author} {\bibfnamefont {Mark~B.}\
  \bibnamefont {Wise}},\ }\href {\doibase 10.1017/9781009402125} {\emph
  {\bibinfo {title} {{Heavy quark physics}}}},\ Vol.~\bibinfo {volume} {10}\
  (\bibinfo {year} {2000})\BibitemShut {NoStop}%
\bibitem [{\citenamefont {Gottfried}\ and\ \citenamefont
  {Yan}(2003)}]{gottfried2003quantum}%
  \BibitemOpen
  \bibfield  {author} {\bibinfo {author} {\bibfnamefont {K.}~\bibnamefont
  {Gottfried}}\ and\ \bibinfo {author} {\bibfnamefont {T.M.}\ \bibnamefont
  {Yan}},\ }\href {https://books.google.com/books?id=8gFX-9YcvIYC} {\emph
  {\bibinfo {title} {Quantum Mechanics: Fundamentals}}},\ Graduate Texts in
  Contemporary Physics\ (\bibinfo  {publisher} {Springer New York},\ \bibinfo
  {year} {2003})\BibitemShut {NoStop}%
\bibitem [{\citenamefont {Schiff}(1955)}]{schiff1955quantum}%
  \BibitemOpen
  \bibfield  {author} {\bibinfo {author} {\bibfnamefont {L.I.}\ \bibnamefont
  {Schiff}},\ }\href {https://books.google.com/books?id=7ApRAAAAMAAJ} {\emph
  {\bibinfo {title} {Quantum Mechanics}}},\ International series in pure and
  applied physics\ (\bibinfo  {publisher} {McGraw-Hill},\ \bibinfo {year}
  {1955})\BibitemShut {NoStop}%
\bibitem [{\citenamefont {Herbst}(1974)}]{Herbst1974}%
  \BibitemOpen
  \bibfield  {author} {\bibinfo {author} {\bibfnamefont {I.~W.}\ \bibnamefont
  {Herbst}},\ }\bibfield  {title} {\enquote {\bibinfo {title} {On the
  connectedness structure of the {Coulomb} {S}-matrix},}\ }\href {\doibase
  10.1007/BF01646192} {\bibfield  {journal} {\bibinfo  {journal}
  {Communications in Mathematical Physics}\ }\textbf {\bibinfo {volume} {35}},\
  \bibinfo {pages} {181--191} (\bibinfo {year} {1974})}\BibitemShut {NoStop}%
\bibitem [{\citenamefont {Lin}(2000)}]{Lin:2000tin}%
  \BibitemOpen
  \bibfield  {author} {\bibinfo {author} {\bibfnamefont {Qiong-gui}\
  \bibnamefont {Lin}},\ }\bibfield  {title} {\enquote {\bibinfo {title} {{On
  the partial wave amplitude of Coulomb scattering in three dimensions}},}\
  }\href {\doibase 10.1119/1.1286117} {\bibfield  {journal} {\bibinfo
  {journal} {Am. J. Phys.}\ }\textbf {\bibinfo {volume} {68}},\ \bibinfo
  {pages} {1056--1057} (\bibinfo {year} {2000})},\ \Eprint
  {http://arxiv.org/abs/quant-ph/0010078} {arXiv:quant-ph/0010078} \BibitemShut
  {NoStop}%
\bibitem [{\citenamefont {Hulth\'{e}n}\ and\ \citenamefont
  {Laurikainen}(1951)}]{HulthenLaurikainen1951}%
  \BibitemOpen
  \bibfield  {author} {\bibinfo {author} {\bibfnamefont {Lamek}\ \bibnamefont
  {Hulth\'{e}n}}\ and\ \bibinfo {author} {\bibfnamefont {K.~V.}\ \bibnamefont
  {Laurikainen}},\ }\bibfield  {title} {\enquote {\bibinfo {title} {Approximate
  eigensolutions of $(d^2\phi/dx^2) + [a + b(e^{-x}/x)]\phi = 0$},}\ }\href
  {\doibase 10.1103/RevModPhys.23.1} {\bibfield  {journal} {\bibinfo  {journal}
  {Rev. Mod. Phys.}\ }\textbf {\bibinfo {volume} {23}},\ \bibinfo {pages}
  {1--9} (\bibinfo {year} {1951})}\BibitemShut {NoStop}%
\bibitem [{\citenamefont {Ma}(1954)}]{Ma:1954}%
  \BibitemOpen
  \bibfield  {author} {\bibinfo {author} {\bibfnamefont {ST}~\bibnamefont
  {Ma}},\ }\bibfield  {title} {\enquote {\bibinfo {title} {On the coulomb and
  hulthen potentials},}\ }\href {\doibase 10.1071/PH540365} {\bibfield
  {journal} {\bibinfo  {journal} {Australian Journal of Physics}\ }\textbf
  {\bibinfo {volume} {7}},\ \bibinfo {pages} {365--372} (\bibinfo {year}
  {1954})},\ \Eprint
  {http://arxiv.org/abs/https://connectsci.au/ph/article-pdf/7/3/365/1345867/ph540365.pdf}
  {https://connectsci.au/ph/article-pdf/7/3/365/1345867/ph540365.pdf}
  \BibitemShut {NoStop}%
\bibitem [{\citenamefont {Dalitz}(1951)}]{Dalitz:1951ah}%
  \BibitemOpen
  \bibfield  {author} {\bibinfo {author} {\bibfnamefont {R.~H.}\ \bibnamefont
  {Dalitz}},\ }\bibfield  {title} {\enquote {\bibinfo {title} {{On higher Born
  approximations in potential scattering}},}\ }\href {\doibase
  10.1098/rspa.1951.0085} {\bibfield  {journal} {\bibinfo  {journal} {Proc.
  Roy. Soc. Lond. A}\ }\textbf {\bibinfo {volume} {206}},\ \bibinfo {pages}
  {509--520} (\bibinfo {year} {1951})}\BibitemShut {NoStop}%
\bibitem [{\citenamefont {Hill}\ \emph {et~al.}(2013)\citenamefont {Hill},
  \citenamefont {Lee}, \citenamefont {Paz},\ and\ \citenamefont
  {Solon}}]{Hill:2012rh}%
  \BibitemOpen
  \bibfield  {author} {\bibinfo {author} {\bibfnamefont {Richard~J.}\
  \bibnamefont {Hill}}, \bibinfo {author} {\bibfnamefont {Gabriel}\
  \bibnamefont {Lee}}, \bibinfo {author} {\bibfnamefont {Gil}\ \bibnamefont
  {Paz}}, \ and\ \bibinfo {author} {\bibfnamefont {Mikhail~P.}\ \bibnamefont
  {Solon}},\ }\bibfield  {title} {\enquote {\bibinfo {title} {{NRQED Lagrangian
  at order $1/M^4$}},}\ }\href {\doibase 10.1103/PhysRevD.87.053017} {\bibfield
   {journal} {\bibinfo  {journal} {Phys. Rev. D}\ }\textbf {\bibinfo {volume}
  {87}},\ \bibinfo {pages} {053017} (\bibinfo {year} {2013})},\ \Eprint
  {http://arxiv.org/abs/1212.4508} {arXiv:1212.4508 [hep-ph]} \BibitemShut
  {NoStop}%
\bibitem [{\citenamefont {Navas}\ \emph {et~al.}(2024)\citenamefont {Navas}
  \emph {et~al.}}]{ParticleDataGroup:2024cfk}%
  \BibitemOpen
  \bibfield  {author} {\bibinfo {author} {\bibfnamefont {S.}~\bibnamefont
  {Navas}} \emph {et~al.} (\bibinfo {collaboration} {Particle Data Group}),\
  }\bibfield  {title} {\enquote {\bibinfo {title} {{Review of particle
  physics}},}\ }\href {\doibase 10.1103/PhysRevD.110.030001} {\bibfield
  {journal} {\bibinfo  {journal} {Phys. Rev. D}\ }\textbf {\bibinfo {volume}
  {110}},\ \bibinfo {pages} {030001} (\bibinfo {year} {2024})}\BibitemShut
  {NoStop}%
\bibitem [{\citenamefont {Hulth\'{e}n}(1942)}]{Hulthen1942}%
  \BibitemOpen
  \bibfield  {author} {\bibinfo {author} {\bibfnamefont {L.}~\bibnamefont
  {Hulth\'{e}n}},\ }\bibfield  {title} {\enquote {\bibinfo {title} {On the
  characteristic solutions of the schr\"odinger deuteron equation},}\
  }\href@noop {} {\bibfield  {journal} {\bibinfo  {journal} {Arkiv f\"{o}r
  Matematik, Astronomi och Fysik}\ }\textbf {\bibinfo {volume} {28A}},\
  \bibinfo {pages} {5--35} (\bibinfo {year} {1942})}\BibitemShut {NoStop}%
\bibitem [{\citenamefont {Jantzen}(2011)}]{Jantzen:2011nz}%
  \BibitemOpen
  \bibfield  {author} {\bibinfo {author} {\bibfnamefont {Bernd}\ \bibnamefont
  {Jantzen}},\ }\bibfield  {title} {\enquote {\bibinfo {title} {{Foundation and
  generalization of the expansion by regions}},}\ }\href {\doibase
  10.1007/JHEP12(2011)076} {\bibfield  {journal} {\bibinfo  {journal} {JHEP}\
  }\textbf {\bibinfo {volume} {12}},\ \bibinfo {pages} {076} (\bibinfo {year}
  {2011})},\ \Eprint {http://arxiv.org/abs/1111.2589} {arXiv:1111.2589
  [hep-ph]} \BibitemShut {NoStop}%
\bibitem [{\citenamefont {Weinberg}(2013)}]{weinberg2013lectures}%
  \BibitemOpen
  \bibfield  {author} {\bibinfo {author} {\bibfnamefont {S.}~\bibnamefont
  {Weinberg}},\ }\href {https://books.google.ch/books?id=WfTq2W_LBlEC} {\emph
  {\bibinfo {title} {Lectures on Quantum Mechanics}}}\ (\bibinfo  {publisher}
  {Cambridge University Press},\ \bibinfo {year} {2013})\BibitemShut {NoStop}%
\bibitem [{\citenamefont {Furry}(1951)}]{Furry:1951bef}%
  \BibitemOpen
  \bibfield  {author} {\bibinfo {author} {\bibfnamefont {W.~H.}\ \bibnamefont
  {Furry}},\ }\bibfield  {title} {\enquote {\bibinfo {title} {{On Bound States
  and Scattering in Positron Theory}},}\ }\href {\doibase
  10.1103/PhysRev.81.115} {\bibfield  {journal} {\bibinfo  {journal} {Phys.
  Rev.}\ }\textbf {\bibinfo {volume} {81}},\ \bibinfo {pages} {115} (\bibinfo
  {year} {1951})}\BibitemShut {NoStop}%
\bibitem [{\citenamefont {Schweber}\ and\ \citenamefont
  {Bethe}(2005)}]{schweber2005introduction}%
  \BibitemOpen
  \bibfield  {author} {\bibinfo {author} {\bibfnamefont {S.S.}\ \bibnamefont
  {Schweber}}\ and\ \bibinfo {author} {\bibfnamefont {H.A.}\ \bibnamefont
  {Bethe}},\ }\href {https://books.google.ch/books?id=MrEqAwAAQBAJ} {\emph
  {\bibinfo {title} {An Introduction to Relativistic Quantum Field Theory}}},\
  Dover Books on Physics\ (\bibinfo  {publisher} {Dover Publications},\
  \bibinfo {year} {2005})\BibitemShut {NoStop}%
\end{thebibliography}%

\end{document}